\documentclass[submission, Phys]{SciPost}
\usepackage{amsmath,amssymb,amsfonts,amsthm}
\usepackage{psfrag,latexsym}
\usepackage{amssymb}
\usepackage{mathtools}
\usepackage{physics}
\usepackage{cancel} 
\usepackage{bm}
\usepackage{dcolumn}
\usepackage{epic} 
\usepackage{epsfig}
\usepackage{feynmp}
\usepackage{graphicx}
\usepackage{grffile}
\usepackage{bm}
\usepackage[usenames, dvipsnames]{xcolor} 
\usepackage{caption}
\usepackage{subcaption}
\usepackage{soul}
\usepackage{wrapfig}
\usepackage{xy} 
\usepackage{xcolor}
\usepackage{tikz}

\usepackage{makecell}

\usepackage{cite}
\usepackage{textcomp}
\usepackage{bbold}
\usepackage[scr=boondox,  
            cal=boondoxo]   
           {mathalpha}
\usepackage[normalem]{ulem}
\usepackage[mathscr]{eucal}

\usepackage{etoolbox}

\makeatletter
\newrobustcmd{\fixappendix}{%
  \patchcmd{\l@section}{1.5em}{7em}{}{}%
  \patchcmd{\l@subsection}{2.3em}{7em}{}{}%
}
\makeatother

\usepackage{color}
\usepackage{dsfont}

\definecolor{mc}{rgb}{0.9,0.3,0.2}
\definecolor{dgreen}{rgb}{0,0.7,0}

\newcommand{\mc}[1]{\mathcal{#1}}  
\newcommand{\msc}[1]{\mathscr{#1}}

\hypersetup{
    colorlinks=true,
    linkcolor=[rgb]{0.55,0,0},
    filecolor=magenta,      
    urlcolor=blue,
    citecolor=blue
}
\usepackage{hyperref}

\begin{document}

\begin{center}{\Large \textbf{
Stochastic dynamics of quasiparticles in the hard rod gas}}\end{center}

\begin{center}
Seema Chahal\textsuperscript{1*},
Indranil Mukherjee\textsuperscript{1$^\dagger$},
Abhishek Dhar\textsuperscript{1$^\ddagger$},
Herbert Spohn\textsuperscript{2$^\S$},\\
Anupam Kundu\textsuperscript{1$^\#$}
\end{center}

\begin{center}
{\bf 1} International Centre for Theoretical Sciences, TIFR, Bengaluru -- 560089, India
\\
{\bf 2} Mathematik Department and Physik Department, Technische Universit{\"a}t\\ M{\"u}nchen, Garching 85748, Germany
\\
\vspace*{0.4 cm}

* seema.s@icts.res.in, 
$\dagger$ indranil.mukherjee@icts.res.in,
$\ddagger$ abhishek.dhar@icts.res.in,\\
$\S$ spohn@ma.tum.de,
$\#$ anupam.kundu@icts.res.in
\end{center}

\begin{center}
\today
\end{center}
\section*{Abstract}
{\bf We consider a one-dimensional gas of hard rods, one of the simplest examples of an interacting integrable model. It is well known that the hydrodynamics of such integrable models can be understood by viewing the system as a gas of quasiparticles. Here, we explore the dynamics of individual quasiparticles for a variety of initial conditions of the background gas. The mean, variance, and two-time correlations are computed exactly and lead to a picture of quasiparticles as drifting Brownian particles. For the case of a homogeneous background, we show that the motion of two tagged quasiparticles is strongly correlated, and they move like a rigid rod at late times. Apart from a microscopic derivation based on the mapping to point particles, we provide an alternate derivation which emphasizes that quasiparticle fluctuations are related to initial phase-space fluctuations, which are carried over in time by Euler scale dynamics. For the homogeneous state, we use the Brownian motion picture to develop a Dean-Kawasaki-type fluctuating hydrodynamic theory, formally having the same structure as that derived recently by Ferrari and Olla~\cite{ferrari2023macroscopic}. We discuss differences with existing proposals on the hydrodynamics of hard rods and some puzzles. }

\vspace{10pt}
\noindent\rule{\textwidth}{1pt}
\tableofcontents\thispagestyle{fancy}
\noindent\rule{\textwidth}{1pt}
\vspace{10pt}

\section{Introduction}
Classical interacting integrable systems in one dimension can be described as a gas of interacting quasiparticles, each tagged by its bare velocity \cite{doyonlecturenotes}. This perspective has proven to be extremely useful for studying the large-scale behavior, enabling the formulation of hydrodynamic equations in terms of the one-particle phase-space density of these quasiparticles, as discussed early on in Ref.~\cite{percus1969exact}. In the last decade, a unified theory known as generalized hydrodynamics (GHD) has been developed for many-body integrable systems, both classical and quantum. The GHD equation takes the form of a collisionless Boltzmann equation written in terms of the phase-space density of quasiparticles. From these equations, one obtains the evolution equations of conserved densities at the ballistic space-time scales.

\begin{figure}[t]
\centering
\includegraphics[width=7.5cm, height=5.6cm]{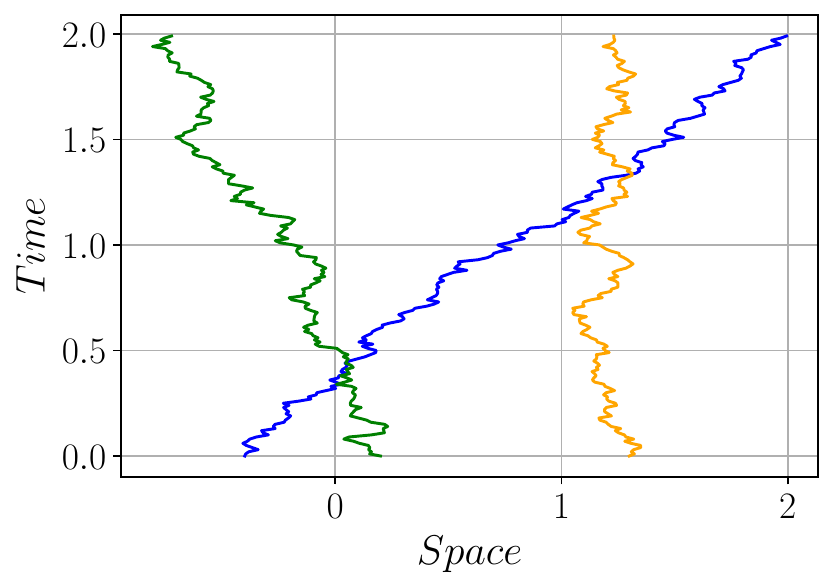}
\vspace*{-0.2 cm}
\caption{Schematic trajectories of three quasiparticles in a one-dimensional integrable system.}
\label{Fig:qp-corr}
\end{figure}

Although the collective motion of quasiparticles successfully describes the large-scale evolution of macroscopic observables, the motion of tagged quasiparticles at the microscopic scale also presents intriguing features. On average, a tagged quasiparticle moves ballistically with an effective velocity~\cite{doyonlecturenotes} arising from collisions with other quasiparticles. The collisions also lead to fluctuations in the quasiparticle's trajectory around the mean ballistic path. This stochasticity can be traced to the initial random configuration of the particles, which leads to irregular collisions with other quasiparticles in both space and time. 
The stochastic motion of quasiparticles has recently been discussed for the case of hard rods in~\cite{ferrari2023macroscopic} and for the Toda system in~\cite{aggarwal2025a,aggarwal2025b}. 
One might expect that two tagged quasiparticles should move independently. However, it turns out~\cite{ferrari2023macroscopic} that they are in fact strongly correlated even at large distances. This happens because the two quasiparticles collide with the same set of background quasiparticles, as illustrated in Fig.~\ref{Fig:qp-corr}. In this paper, we investigate this stochastic motion of tagged quasiparticles in a system of hard rods. 

\begin{figure}[t]
\centering
\includegraphics[width=7.4cm, height=5.5cm]{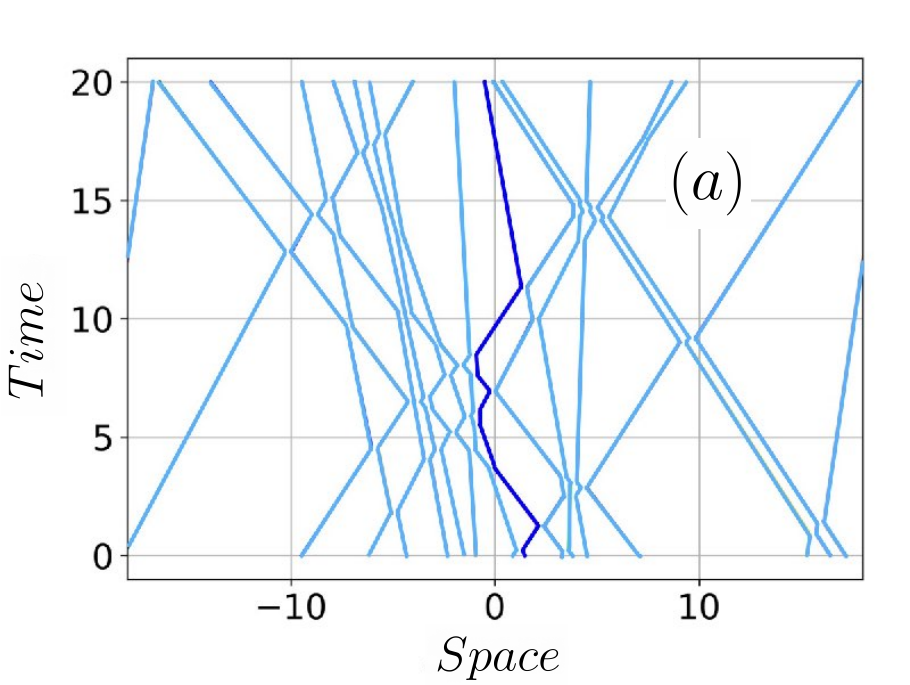}
\includegraphics[width=7.4cm, height=5.5cm]{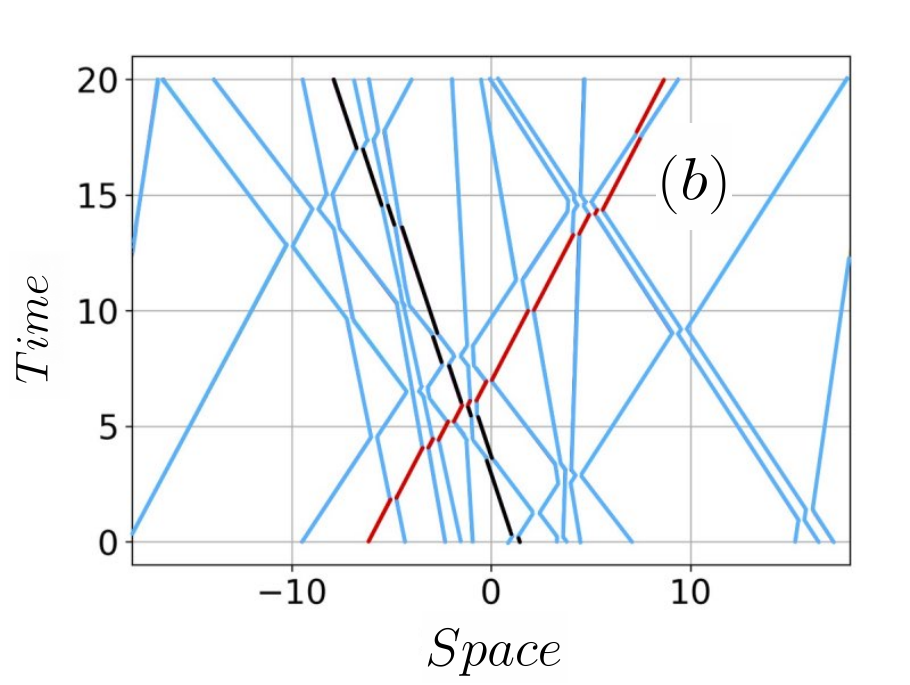}
\vspace*{-.2 cm}
\caption{The schematic diagram (a) illustrates the stochastic trajectories of hard rods encountering collisions and exchanging velocities with each other. The trajectory marked in deep blue indicates the path of a marked rod. In (b), we show trajectories of two quasiparticles (black and red) tagged by their bare velocities.  At each collision, a quasiparticle jumps by a distance $\pm a$ while retaining its velocity. Since the collisions occur at random times for random initial configurations, a quasiparticle follows a stochastic path around a mean ballistic motion with an effective velocity.}
\label{Fig:quasiparticle_def}
\end{figure}

The hard rod gas provides a particularly simple example in which the quasiparticles are tagged by their bare velocities and undergo fixed jumps in position on collisions with other quasiparticles. The motion of the rods and the quasiparticles is illustrated in Fig.~\ref{Fig:quasiparticle_def}. Using a microscopic approach, we characterize the stochastic dynamics of quasiparticles. The diffusion of quasiparticles and correlations between two spatially separated quasiparticles for a gas in a globally homogeneous equilibrium state was recently discussed by Ferrari and Olla using a rigorous probabilistic approach~\cite{ferrari2023macroscopic}. Here, we present a physical derivation of these results and extend them to the case of inhomogeneous initial states of the gas (e.g., a domain-wall configuration). Below we outline our  main contributions along with the plan of the paper.

\begin{enumerate}


\item Section~\ref{sec:2} introduces the microscopic dynamics of hard rods and describes the well-known mapping to the hard-point gas model \cite{percus1969exact,singh2024thermalization,Mrinal_2024_HR}. For ease of computations, we choose our initial conditions in a special way, which we now specify.  To study the single quasiparticle statistics, we place a special rod at the origin with velocity $v_0$ and distribute the background rods on  both sides, with their positions and velocities sampled independently from specified distributions. For the two quasiparticle case, we place two special rods at fixed locations (one at the origin and the other at a fixed distance $\Delta X$) with specified velocities $v_0,~u_0$. The background rods are then distributed randomly in the left, middle (in between the two special rods) and right regions, sampled independently from the given distributions.

\item In Section~\ref{sec:3} we analyze the stochastic motion of a single quasiparticle that starts at the origin with bare velocity $v_0$. We denote the position of the quasiparticle by $X_{v_0}(t)$ at time $t$. We  re-derive the result for the variance of  $X_{v_0}(t)$ of a single quasiparticle obtained previously in \cite{Lebowitz_Percus1967, singh2024thermalization, Mrinal_2024_HR}. We find a diffusive growth of the variance, with a diffusion constant that depends on the bare velocity $v$ and the density of the gas. Next, we present our first new result,  on the position autocorrelation of a quasiparticle $\langle X_{v_0}(t_1)X_{v_0}(t_2)\rangle_c=\langle X_{v_0}(t_1)X_{v_0}(t_2)\rangle-\langle X_{v_0}(t_1)\rangle \langle X_{v_0}(t_2)\rangle$.  We find that the expressions of the autocorrelation have the same dependence on $t_1,t_2$ as a Brownian particle. Although, we can compute these for general inhomogeneous initial conditions, and the expressions simplify for the homogeneous and domain-wall initial conditions.

\item Section~\ref{sec:two_quasiparticles} analyzes the motion of two quasiparticles, starting at positions $X_{v_0}(0)=0$ and $X_{u_0}(0)=\Delta X$  with velocites $v_0$ and $u_0$, respectively. We derive an explicit expression for the correlation $\langle X_{v_0}(t)  X_{u_0}(t)\rangle_c$. For the homogeneous case, our result is in agreement with the results of \cite{ferrari2023macroscopic}. In particular, for $v_0=u_0$ the quasiparticles  become perfectly correlated  even if initially they start far apart ({\it i.e.,} large $\Delta X$) at very large times of $ O(\Delta X^2)$ and in the thermodynamic limit. This makes them move  effectively  like a rigid body. We find that this rigid-body-like behavior persists even when they move in an inhomogeneous background such as an initial domain-wall profile. 

\item The fluctuations and correlations of the tagged quasiparticles arise essentially from initial fluctuations in the phase-space densities that are carried ballistically to time $t$ by Euler GHD. In Section~\ref{corr-frm-EGHD}, we demonstrate this fact by re-deriving the expressions of the variance, autocorrelation, and covariance using space-time correlations of mass densities on the ballistic scale.

\item  In Section~\ref{sec:quasiparticles_fluc_hd}, we consider the description of the hard rod gas as a collection of correlated Brownian quasiparticles. We then use the Dean-Kawasaki~\cite{dean1996langevin,illien2024dean} formalism to present a phenomenological derivation of the fluctuating hydrodynamics equation for this system.

\item Section~\ref{sec:two quasiparticles quenched} extends our study on quasiparticle to quenched initial conditions and finds an explicit expression for the variance and covariance of two quasiparticles.

\end{enumerate} 
We conclude in Section~\ref{sec:conclusion} by discussing some puzzles and outlining potential directions for future research. Supplementary derivations are provided in the appendix.

\section{Hard rod dynamics and initial conditions}
\label{sec:2}

Let the positions and velocities of the $\msc{N}$ hard rods be represented by $\{X_i\}$ and  $\{V_i\}$ for $i=-N_1,...,0,...,N_2$ with $\msc{N}=N_1+N_2+1$. The positions are ordered in the sense $X_{i+1} \ge X_{i}+a$.   These rods undergo ballistic motion between instantaneous elastic collisions, and during each collision, their velocities (since they each have unit mass) are exchanged. Clearly, in the limit $a=0,$ {\it i.e.} when the rod length approaches zero, the interacting hard-rod system reduces to a non-interacting hard-point gas (HPG).  The microscopic dynamics of hard rods can be mapped onto a system of hard-point particles through a specific transformation. Starting from a configuration of $\msc{N}$ hard rods $\{X_i,V_i\}$ at any instant, one can construct a configuration of $\msc{N}$ hard-point particles $\{x_i,v_i\}$ (each of unit mass) by the transformation \cite{percus1969exact,bernstein1988expansion,lebowitz1968time}.
\begin{equation}
\label{eq:map-hr-hp}
    x_i =X_i - ia,~~ v_i = V_i, ~~\text{for} ~ i=-N_1,\dots, N_2.
\end{equation}
The above transformation essentially is obtained by excluding the inaccessible space between successive rods. In the point particle representation, the dynamics become simple. They move ballistically, and at collisions, they just exchange their velocities without suffering from any jump in the position, unlike the hard rods. Hence, one can just evolve them ballistically as non-interacting particles and at the final time relabel them according to the order of their positions. Since the mapping in Eq.~\eqref{eq:map-hr-hp} is one-to-one, one can also transform to hard rods back from hard-point particles. Hence, using this mapping, the hard rod dynamics can be fully solved. 

In this paper, we consider a certain class of initial conditions. The initial configurations are first chosen in hard-point coordinates and then transformed to hard rod coordinates. We first choose $\msc{N}=N_1+N_2+1$ locations $\{\hat{x}_i\}$ for the point particles inside a domain $[\msc{L}_1,\msc{L}_2]$ of size $\msc{L}=|\msc{L}_2-\msc{L}_1|$. We choose these locations independently and identically from a distribution $\mc{p}_a(\hat{x})$ such that $\int_{\msc{L}_1}^{\msc{L}_2} dx~\mc{p}_a(x)=1$. We then arrange the locations in increasing order to get the coordinates: $\{x_i\}=\text{Order}[\{\hat{x}_i\}]=\{x_i~;~\msc{L}_1<x_{-N_1}<...<x_{N_2}<\msc{L}_2\}$ of the $\msc{N}$ hard-point particles. Once we have the ordered locations $\{x_i\}$ of the particles, for each $i$, we choose a velocity independently of the distribution $\mc{h}(u_i)$ such that $\int _{-\infty}^\infty\mc{h}(u)du=1.$ The particles are then allowed to move throughout the space.  
The joint distributions of the positions and velocities of the $\msc{N}$ hard-point particles is given by
 \begin{align}
 \mathbb{P}_{a}(\{x_{i},u_{i}\},0)=\msc{N}!\prod_{i=-N_1}^{N_2} \mc{p}_{a}(x_{i})\mc{h}(u_{i})~\prod_{i=-N_1}^{N_2-1}\Theta(x_{i+1}-x_i)
 .\label{en-aa}
 \end{align}
The product over the Heaviside functions $\Theta$ ensures the ordering $\{x_{i}\le x_{i+1}~;~i=-N_1,...,N_2\}$.
We eventually consider the thermodynamic limit $\msc{N} \to \infty$ and $\msc{L} \to \infty$ such that the initial mass density profile becomes a finite-valued function $\bar{\varphi}(x)$ everywhere.
 The initial mean phase space density (PSD) and the mean mass density of the hard-point particles are given by $\bar{f}(x,u) = \msc{N} \mc{p}_a(x)\mc{h}(u)$ and  $\bar{\varphi}(x)=\msc{N}\mc{p}_a(x)$, respectively. For each configuration $\{x_i,v_i\}$ of the positions and velocities of the point particles, we construct a configuration of the hard rods using the inverse mapping of Eq.~\eqref{eq:map-hr-hp}
 \begin{align}
 X_i = x_i + ia,~~\text{and}~~V_i=v_i~~\text{for}~i=-N_1,...,N_2. 
  \label{inv-map-to-hpg-line}   
 \end{align}
Thus, we have an ensemble of hard-rod configurations characterized by the average mass density profile $\bar{\varrho}(X(x)) = \frac{\bar{\varphi}(x)}{1+a\bar{\varphi}(x)}$ with $X(x)=x+a\int_{0}^x dy~\bar{\varphi}(y)$. Note that the transformation to the hard rod position coordinate $X(x)$ from the hard-point position coordinate $x$ is essentially a restatement of the above mapping in Eq.~\eqref{inv-map-to-hpg-line} in terms of the mass density $\bar{\varphi}(x)$ of the point particles.
Also note that after the coordinate transformation \eqref{inv-map-to-hpg-line} the densities of hard rods get correlated over a long distance,except for the case of homogeneous or domain wall initial conditions. This is in contrast to the local Generalized Gibbs initial state of hard rods \cite{doyon2017dynamics,doyon2023ballistic, doyonlecturenotes}, where the correlation between hard rod densities  decays exponentially with separation.
Since the velocities of the point particles and the hard rods remain the same under the mapping in Eqs.~\eqref{eq:map-hr-hp} and \eqref{inv-map-to-hpg-line}, from now on we will represent the velocities of the rods by $v_i$ instead of $V_i$ for $i=-N_1,...,N_2$. For all our numerical simulations, we chose $\mc{h}(u)$ to be a mean-zero Maxwell distribution at temperature $T$.

\section{Dynamics of a single tagged quasiparticle}
\label{sec:3}

We consider an initial configuration of $\msc{N}=2N+1$ hard rods on a one-dimensional line, where the quasiparticle with velocity $v_0$ is positioned at the origin, with $N$ rods placed to its left over the region $[-L,0]$ and $N$ rods to its right {\it i.e.,} over the region $[0,L]$. To choose such a hard rod configuration, we follow the procedure given in the previous section [see Eq.~\eqref{en-aa}]. We choose positions of $N$ point particles each on both sides of the tagged point particle at the origin randomly and independently sampled from the distributions $\mc{p}_\ell({x})=\frac{\varphi_\ell({x})}{N}$ and  $\mc{p}_r({x})=\frac{\varphi_r({x})}{N}$, respectively, such that the mean initial mass density becomes $\bar{\varphi}(x)= \varphi_\ell({x})\Theta(-x) + \varphi_r({x})\Theta(x)$. The velocities of all the point particles, except for the quasiparticle, are independently sampled from the velocity distribution $\mc{h}(v)$. By indexing the point particles as $\{x_i,v_i;i=-N,...,N\}$, we use the inverse mapping in Eq.~\eqref{inv-map-to-hpg-line} to get the corresponding configuration of the hard rods.

The displacement at time $t$ of a quasiparticle, starting from the origin and  with velocity tag $v_0$,  is given by 
\begin{equation}\label{eq:1quasiparticle_Xt}
X(t)=v_0 t+a[n_{r\ell}(t)-n_{\ell r}(t)],    
\end{equation}
where $n_{r\ell}(t)$ is the number of rods that collided with the quasiparticle from the right and $n_{\ell r} (t)$ is the number of rods that collided from the left up to time $t$. The numbers $n_{r\ell}(t)$ and $n_{\ell r}(t)$ are random as they fluctuate between different initial conditions. The statistics of $X(t)$ is thus completely determined by those of  $n_{r\ell}(t)$ and $n_{\ell r}(t)$, which are independent random variables and whose distributions are easy to obtain by using the mapping to hard-point particles in Eq.~\eqref{eq:map-hr-hp}.

Thus, in order to find statistics of $X(t)$, we need the distributions  $\mathcal{P}_R,\mathcal{P}_L$  of $n_{r\ell}(t)$ and $n_{\ell r}(t)$ respectively. To determine these distributions, we simply need the probability that, in time $t$, a background point particle crosses the trajectory of the corresponding tagged point particle that reaches position $x(t)=v_0t$ starting from the origin. Let us denote the probabilities of crossing from right by $p_{r \ell}/N$  and from left by $p_{\ell r}/N$, where the explicit expressions of $p_{r\ell}$ and $p_{\ell r}$ are given in Appendix \ref{sec:appA} [see Eq.~\eqref{eq:1quasiparticle_part-probs}]. In terms of these, it is easy to see~\cite{Mrinal_2024_HR} that $\mathcal{P}_R,\mathcal{P}_L$ are binomial distributions of the form
\begin{align}
\mathcal{P}(n,t) = \binom{N}{n}~ \left(\frac{p}{N}\right)^n \left(1-\frac{p}{N}\right)^{(N-n)}~,
\label{eq:binom} 
\end{align}
where $p=p_{r \ell}$ for $\mathcal{P}_R$ and $p=p_{\ell r}$ for $\mathcal{P}_L$. For large $N$, we get Poisson-distributed number fluctuations:
\begin{align}
\mathcal{P}_R(n,t) = \frac{p_{r \ell}^n}{n!}e^{-p_{r \ell}},~~\mathcal{P}_L(n,t) = \frac{p_{\ell r}^n}{n!}e^{-p_{\ell r}}.
\label{eq:binom} 
\end{align}
We can immediately write the mean and variances:
\begin{subequations}
\label{mean-var-Z}
 \begin{align}
    \langle X (t)\rangle &= v_0t +a [p_{r \ell}(t)-p_{\ell r}(t)], \label{mean-Z}\\ 
    &  \notag \\ 
    \langle X^2(t)\rangle_c &= a^2 [p_{\ell r}(t)+p_{r \ell}(t)], \label{var-Z}
\end{align}
\end{subequations}
where we have used $\langle n_{r\ell}(t)\rangle=\langle n^2_{r\ell}(t)\rangle_c=p_{r\ell}(t)$ and $\langle n_{\ell r}(t)\rangle=\langle n^2_{\ell r}(t)\rangle_c=p_{\ell r}(t)$.
The full distribution of $X(t)$ can be  obtained~\cite{Mrinal_2024_HR}
from the corresponding generating function,
\begin{align}
\langle e^{\mc{i} k X(t)}\rangle = e^{\mc{i}  k v_0 t} e^{p_{r \ell}(e^{\mc{i} k{ a}}-1)}e^{p_{\ell r}(e^{-\mc{i}  k{ a}}-1)}. \label{Xgenf}
\end{align}
For certain initial conditions, such as homogeneous or domain wall initial conditions, the functions $p_{r\ell}(t)$ and $p_{\ell r}(t)$ would grow linearly with time and hence all  cumulants would also have the same scaling, consequently we would have a large deviation form. At large times, the typical distribution of $X(t)$ is Gaussian
\begin{equation}\label{eq:P(Z)-gauss}
    \mathbb{P}(X,t) \approx  \frac{1}{\sqrt{2 \pi \langle X(t) ^2\rangle_c}}~\exp\left( - \frac{(X - \langle X(t) \rangle)^2}{2 \langle X(t)^2\rangle_c}\right),
\end{equation}
for $X-\langle X(t) \rangle \lesssim \sqrt{\langle X(t)^2 \rangle_c}$.

We now consider the two-time correlation $\langle X(t_1) X(t_2) \rangle_c$. It is clear from Eq.~\eqref{eq:1quasiparticle_Xt} that they depend on the time correlation of number fluctuations as
\begin{align}
    \langle X(t_1) X(t_2) \rangle_c = a^2 \left[\langle n_{r \ell}(t_1) n_{r \ell}(t_2) \rangle_c +\langle n_{\ell r}(t_1) n_{\ell r}(t_2) \rangle_c \right].
\end{align}
To evaluate the correlation, $\langle n_{r \ell}(t_1) n_{r \ell}(t_2) \rangle$, we need the joint probability that $n_1$ hard-point particles crossed the tagged particle in time $t_1$ from the right and $n_2$ particles crossed it in time $t_2$. Now we note that, assuming $t_2> t_1$, the joint probability distribution ${\rm Prob}(n_1,t_1; n_2, t_2)$ is simply given by $\mathcal{P}_R(n_1,t_1) \mathcal{P}_R(n_2-n_1,t_2-t_1) $. Hence $\langle n(t_1) n(t_2)\rangle_c = \langle n^2(t_1) \rangle$  and  then it follows that, for $t_1 < t_2$, 
\begin{align}
   \langle X(t_1) X(t_2) \rangle_c = a^2 ( \langle n_{r \ell}^2(t_1)\rangle + \langle n_{\ell r}^2(t_1)\rangle)=\langle X^2(t_1)\rangle_c = a^2 [p_{\ell r}(t_1)+p_{r \ell}(t_1)].  
   \label{1t:acorr}
\end{align}
We now state the explicit forms of $\langle X(t) \rangle$ and $\langle X^2(t)\rangle_c$ for two choices of the distributions of the background particles. 
\begin{itemize}
    \item{Homogeneous case:} For the special case of a homogeneous initial distribution, i.e,
    $\varphi_\ell=\varphi_r=\varphi_0$, the expressions of $p_{\ell r}$ and $p_{r\ell}$ can be obtained easily. In the limit of $N \to \infty$, the quasiparticle simply sees a constant flux of particles, given by
$(w-v_{0})\varphi_0 $ from the left and $(v_{0}-w) \varphi_0$ from the right. Hence, the mean number of particles crossing the quasiparticle from the left and right respectively, are given by,
\begin{align}
\label{eq:hom-pform}
p_{\ell r} = t \int_{v_0}^\infty dw ~ \varphi_0 \mc{h}(w) (w-v_0), ~ p_{r \ell} = t \int_{-\infty}^{v_0} dw ~ \varphi_0 \mc{h} (w) (v_0-w).
\end{align} 
 Using these in 
    Eqs.~\eqref{mean-var-Z} and \eqref{1t:acorr} gives us: 
\begin{subequations}
\begin{align}
 \langle X (t)\rangle &= v_{\rm eff} t,~~~\langle X^2(t)\rangle_c =  \mathcal{D}(v_0) t,~~\text{and} \label{mean-var-Z-homo-a}\\
  \langle X(t_1)&X(t_2)\rangle_c =\mc{D}(v_0)\min(t_1,t_2),\label{eq:2time_corr_1quasiparticle_expression} \\
 {\rm where}~~v_{\rm eff}&= \frac{v_0-a \varrho_0 \mc{u}}{1-a \varrho_0},~~\mathcal{D}(v_0)= a^2 \varphi_0 \int dw |v_0-w| \mc{h}(w), 
 \label{mean-var-Z-homo-b}
\end{align} 
 \label{mean-var-Z-homo}
\end{subequations}
with $\varrho_0 = \frac{\varphi_0}{1+a\varphi_0}$ and $\mc{u}=\int_{-\infty}^\infty dw ~w \mc{h}(w)$. 
In Fig.\ref{Fig:1quasiparticle_origin_time_corr}a, we provide the numerical verification of the above expression of the two-time correlation function.

\begin{figure}[t]
\centering
\includegraphics[width=6.9cm, height=5.5cm]{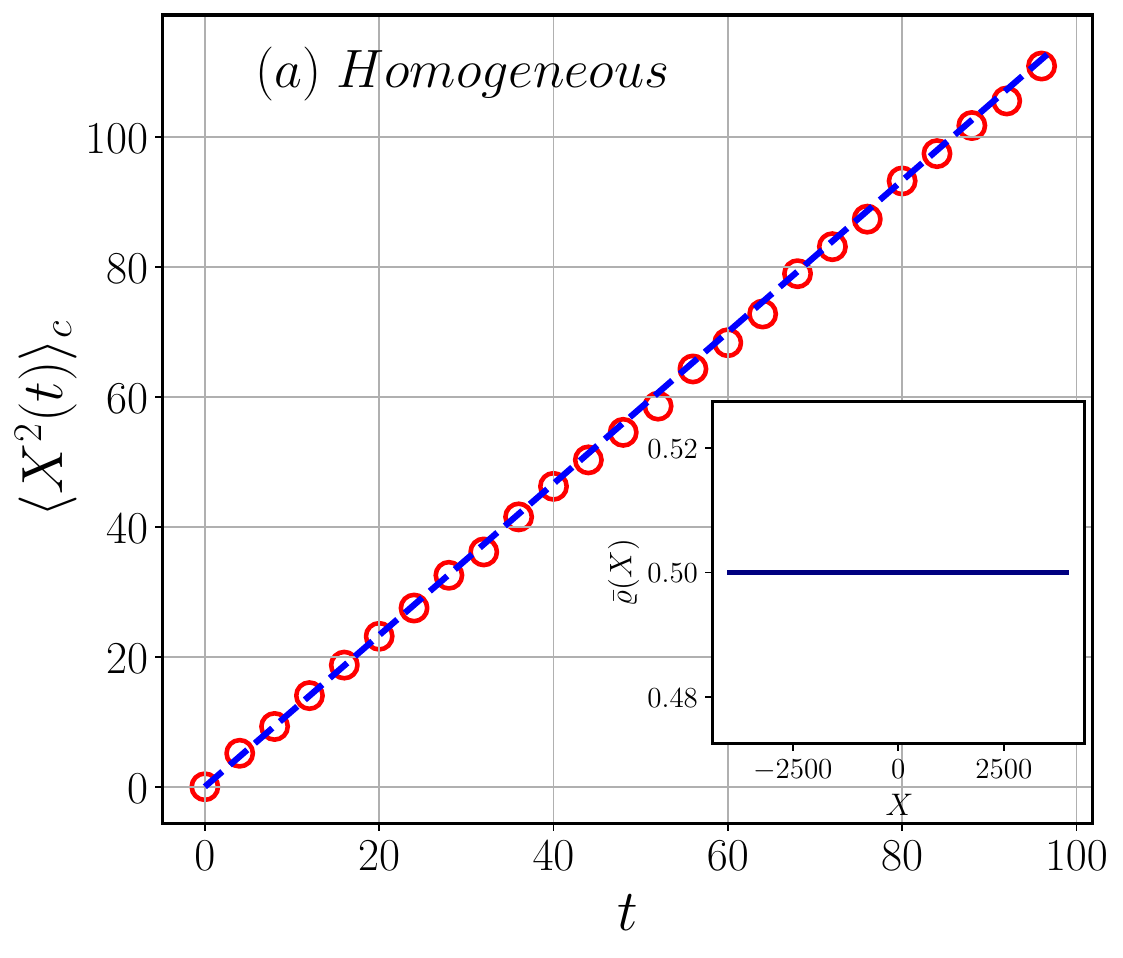}
\includegraphics[width=7.0cm, height=5.5cm]{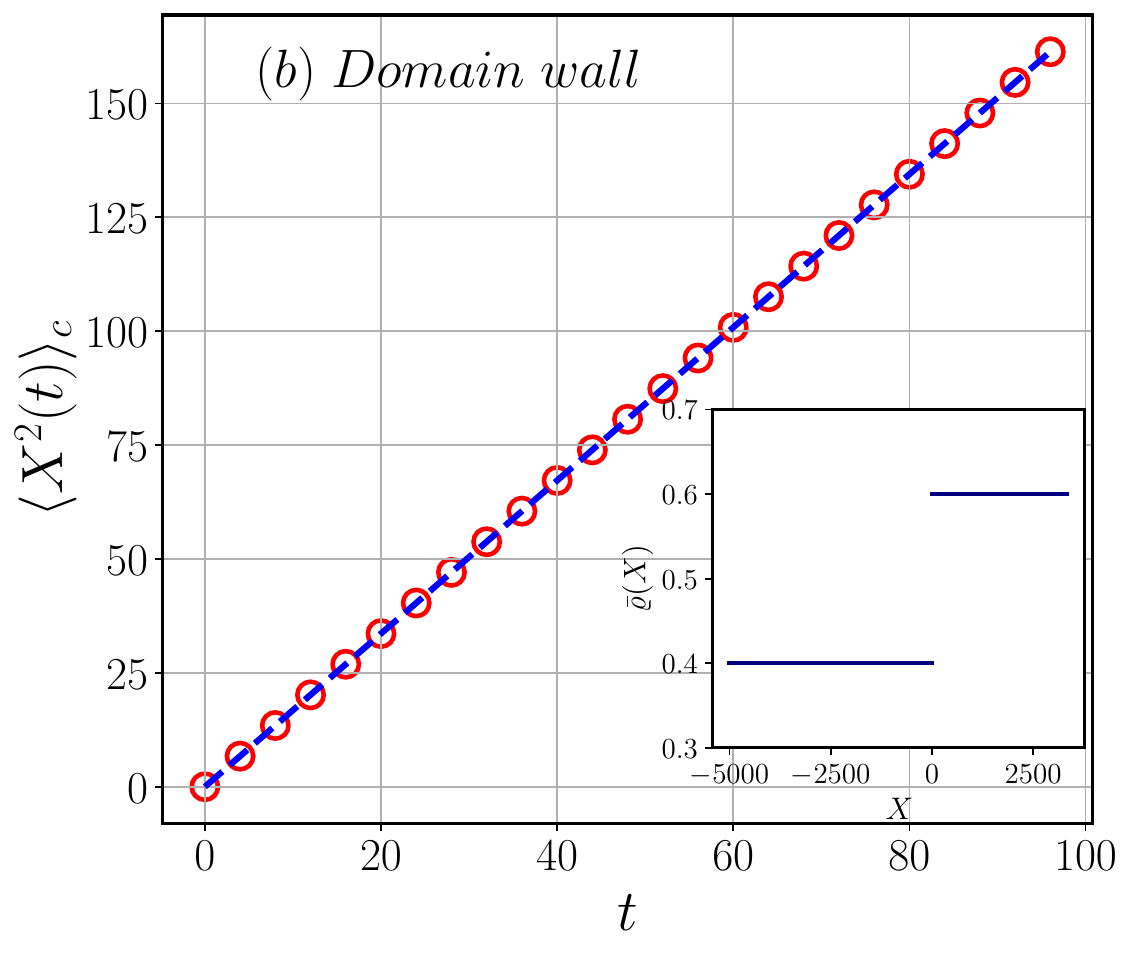}
\includegraphics[width=6.9cm, height=5.5cm]{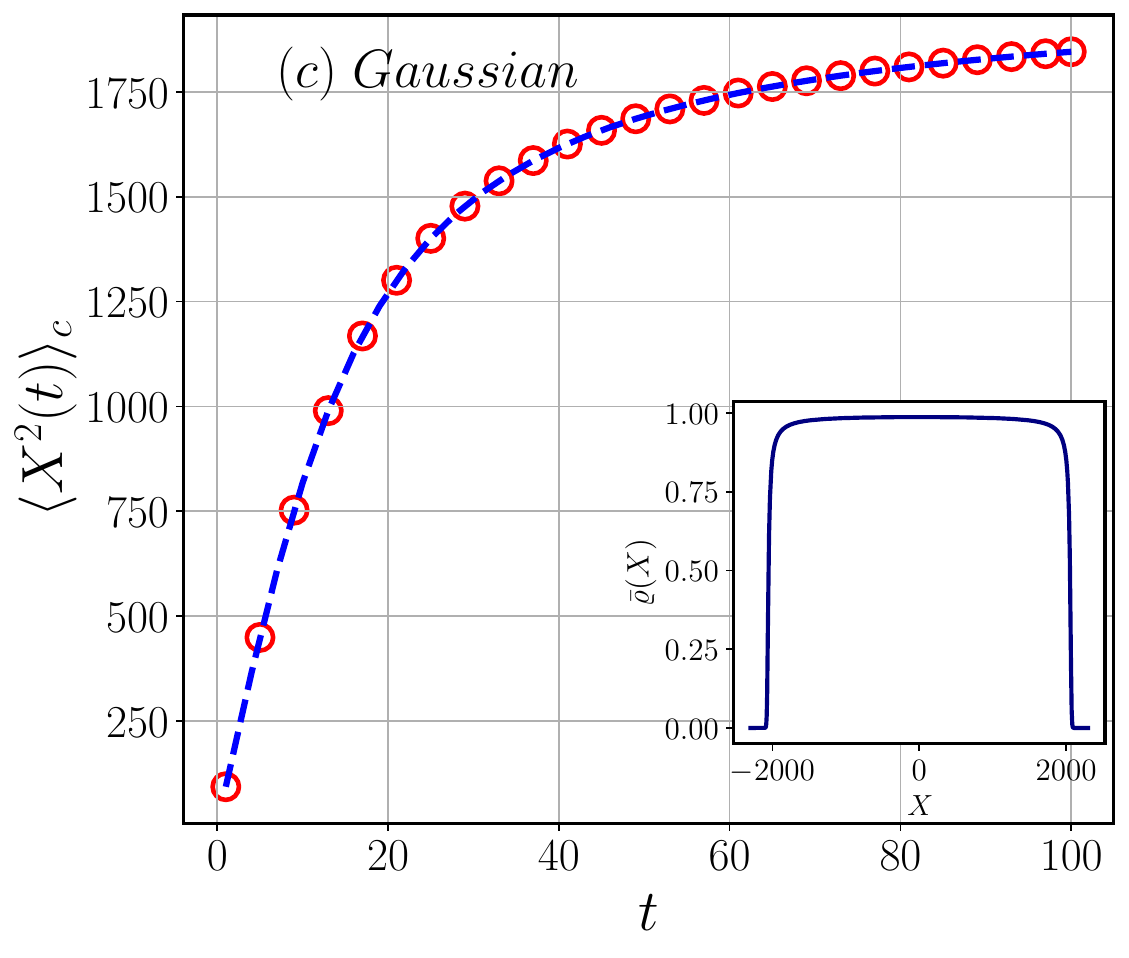}
\includegraphics[width=7.0cm, height=5.5cm]{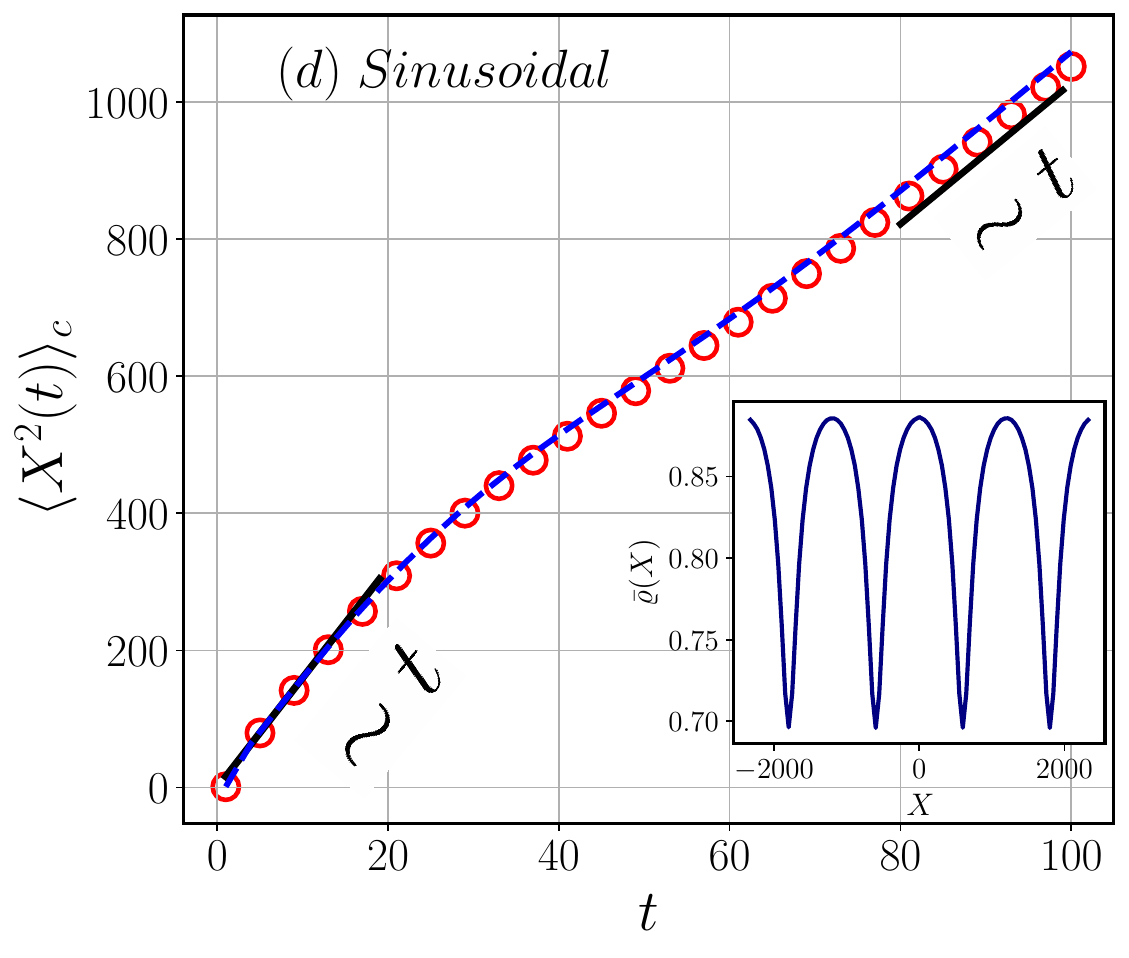}
\vspace*{-.2 cm}
\caption{This figure compares the numerical simulation (circles) with the exact results (dashed line) of the variance of a single quasiparticle moving with velocity $v_0=1$ for different initial density profiles (see insets): (a) homogeneous profile, (b) domain-wall profile,
(c) Gaussian profile, and (d) sinusoidal density.  
In all cases $N=2000$, $a=1.0$ and $T=1.0$.}
\label{Fig:1quasiparticle_origin_time_corr}
\end{figure}

\item{Domain wall case:} For the special case of domain-wall initial condition with $\bar{\varphi}(x)=\varphi_\ell \Theta(-x)+\varphi_r\Theta(x)$, we can repeat the arguments for the homogeneous case above, to obtain the expressions of $p_{\ell r}$ and $p_{r\ell}$ which are 
\begin{align}
\label{eq:pform}
p_{\ell r} = t \int_{v_0}^\infty dw ~ \varphi_\ell \mc{h}(w) (w-v_0), ~ p_{r \ell} = t \int_{-\infty}^{v_0} dw ~ \varphi_r \mc{h} (w) (v_0-w).
\end{align}
Inserting these forms in Eq.~\eqref{mean-var-Z} and defining $\varphi_0 = \frac{\varphi_\ell+\varphi_r}{2}$ and $\delta \varphi = \frac{\varphi_\ell-\varphi_r}{2}$, one gets 
\begin{align}
 \langle X (t)\rangle &= v^{\rm dw}_{\rm eff} t, ~
 \langle X^2(t)\rangle_c =  \mathcal{D}_{\rm dw}(v_0) t, ~\text{and}~  \langle X(t_1)X(t_2)\rangle_c =\mc{D}_{\rm dw}(v_0)\min(t_1,t_2),
 \label{mean-var-Z-dw-a}
 \end{align}
 where
 \begin{align}
 \begin{split}
v^{\rm dw}_{\rm eff}&= \frac{v_{ 0}-a \varrho_0 \mc{u}}{1-a \varrho_0}+a \delta \varphi \int dw |v_{ 0}-w| \mc{h}(w), \\ 
 \mathcal{D}_{\rm dw}(v_0)&= a^2 \varphi_0\int dw |v_0-w| \mc{h}(w) - a^2 \delta \varphi  (v_0 - \mc{u}), 
 \end{split}
  \label{mean-var-Z-dw-b}
\end{align}
with $\varrho_0 = \frac{\varphi_0}{1+a\varphi_0}$. It is interesting to note that in the case of a domain-wall initial condition, both the effective speed and the diffusion constant change from their values in the homogeneous case [given in Eq.~\eqref{mean-var-Z-homo}], and the changes in both cases are proportional to the inhomogeneity $\delta \varphi$.

\item Other inhomogeneous initial conditions: Note the expression of the mean, variance, and autocorrelation in Eqs.~\eqref{mean-var-Z} and \eqref{1t:acorr} are valid for initial conditions with arbitrary mass density and velocity distributions. They become 
\begin{align}
\begin{split}
\langle X_{v_0}(t) \rangle &= a \int_0^t dt' \int_{-\infty}^\infty dv~ (v-v_0)~\bar{f}(v_0t',v,t'),  \\ 
\langle X^2_{v_0}(t) \rangle_c &= a^2 \int_0^t dt' \int_{-\infty}^\infty dv~ |v-v_0|~\bar{f}(v_0t',v,t'), 
\end{split}\label{meanX-inhomo} 
\end{align}
where $\bar{f}(x,v,t) $ is the average phase space density of the hard-point particles at time $t$. 
In Fig.~\ref{Fig:1quasiparticle_origin_time_corr}, we plot the variance of the quasiparticle for different initial conditions of the form $ \bar{f}(x,v,0)  =\bar{\varphi}(x) \mc{h}(v)$, within the point particle picture. Panels (a) and (b) correspond to homogeneous and domain-wall initial conditions, respectively. Panel (c) and (d) correspond to a Gaussian initial profile, $\bar{\varphi}(x)=\frac{1}{\sqrt{2\pi \sigma^2}}e^{-\frac{x^2}{2\sigma^2}}$ with $\sigma=20$, and a sinusoidal profile, $\bar{\varphi}(x) = 0.9+0.5 \cos(\pi x/100)$, respectively. 
The insets show the plots of the corresponding initial mass density profiles of the hard rods obtained from $\bar{\varphi}(x)$ using $\bar{\varrho}(X(x))=\frac{\bar{\varphi}(x)}{1+a\bar{\varphi}(x)}$ with $X(x)=x+a\int_{0}^x dy~\bar{\varphi}(y)$. We observe that the variance at short times grows linearly with time; however, at late times it saturates for the Gaussian case and changes slope in the sinusoidal case. At short time, the quasiparticle sees the local peak of the initial density profile. As time grows, it starts seeing the inhomogeneity, and at very large time it has moved (ballistically) by a distance comparable to the scale of the inhomogeneity of the density profile. Since in the Gaussian case, the density almost vanishes over this scale, the quasiparticle finds it difficult to further collide with background quasiparticles, and hence the variance does not grow. On the other hand, for the sinusoidal case, at late time, the quasiparticle sees the background with an overall homogeneous density that is different from the initial value at the starting point of the quasiparticle. Consequently, the slope of the variance changes at large times in this case. 

\end{itemize}

\noindent
 It is interesting to note that the correlation function in Eq.~\eqref{eq:2time_corr_1quasiparticle_expression} is directly proportional to the diffusion constant ${\mc{D}}(v_0)$, which is determined by the velocity of the quasiparticle $v_0$ and depends on the minimum of the two times $t_1,t_2$. Notably, this correlation of a single quasiparticle closely resembles that of a free Brownian particle, governed by the stochastic differential equation 
 \begin{align}
   \frac{dX_{v_0}}{dt}=v_{\rm eff}(v_0)+\xi_{v_0}(t),  
 \end{align} 
 where $\xi_{v_0}(t)$ represents Gaussian white noise with $\langle \xi_{v_0}(t_1)\xi_{v_0}(t_2)\rangle=2\mc{D}(v_0)\delta(t_1-t_2)$. This indicates that quasiparticles for a homogeneous and domain-wall background of other rods move effectively as a  Brownian particle -- a fact that was established (for the homogeneous case) in \cite{ferrari2023macroscopic}. We next investigate the correlation between different quasiparticles.

\section{Dynamics of two quasiparticles}\label{sec:two_quasiparticles}
The previous section provided a brief overview of the diffusion of a single quasiparticle, outlining the microscopic approach used to determine its distribution, following the Ref. \cite{Mrinal_2024_HR}. In this section, we extend the analysis to the case of two quasiparticles. 

We label these two quasiparticles as $X(t)$ and $Y(t)$ which are initially positioned at $X(0)=X_0 =0$ and $Y(0)=Y_0>0$ with velocities $v_0$ and $u_0>v_0$, respectively. We place $N$ rods on the left of $X_0$ and $N$ rods on the right of $Y_0$, and $\bar{N}$ number of rods in between. Clearly, $Y_0 \ge (\bar{N}+1)a$. As before, the positions and velocities of the background rods on the left of $X_0$, on the right of $Y_0$, and in the middle are first chosen in the point particle picture following the distribution in Eq.~\eqref{en-aa} and then transformed to hard rod coordinates using the mapping in Eq.~\eqref{inv-map-to-hpg-line}.  Let the statistical state correspond to point particle density $\varphi_\ell(x)$ on the left of $X_0$, $\varphi_r(x)$  on the right of $Y_0$ and $\varphi_m(x)$ in between $X_0$ and $Y_0$. The initial mass density profile of the point particles is given by 
\begin{align}
    \bar{\varphi}(x)=\varphi_\ell(x)\Theta(-x) +\varphi_m(x)\Theta(y_0-x)\Theta(x) +\varphi_r(x)\Theta(x-y_0), \label{ini-den-2tag}
\end{align}
where $y_0=Y_0-(\bar{N}+1)a$.

As the system evolves, the quasiparticles move, and we denote the positions of the two quasiparticles at some later time $t$ by $X(t)$ and $Y(t)$, respectively.  
As we have seen in the last section, each of the two quasiparticles undergoes Brownian motion as a result of collisions with the background rods. Correlations between the motion of the two rods emerge because the two tagged quasiparticles might collide with the same set of background quasiparticles. 
Here, we primarily focus on determining the mean, variance, and covariance of the positions of the two quasiparticles at time $t$.

As already noted, we assume here \( v_0 \le u_0 \), though our computation can be easily extended to the reversed case.  Since, $Y_0>0$,  the quasiparticle $X(t)$ never crosses $Y(t)$.
Let  $n_{r\ell}(t)~(n_{\ell r}(t))$ denote the number of rods that were initially to the right (left) of $Y_0$ ($X_0$) and then collided with both $X(t)$ and $Y(t)$ during the time interval $t$. 
We let $n_{r m}(t)$ ($n_{\ell m}(t)$) denote the number of rods that were initially to the right (left) of $Y_0$ ($X_0$), that collided only with $Y(t)$ ($X(t)$) during time $t$. Finally, we let  $n_{m \ell}(t) (n_{mr}(t))$ denote the number of rods that were initially present between $X_0$ and $Y_0$ and which collided with $X(t) (Y(t))$ during time $t$ (see Fig.~\ref{Fig:2quasiparticle_separated}). 
Then, for any given initial configuration of the background rods, the positions of the quasiparticles at time $t$ are given by
\begin{align}\label{eq:two quasiparticles_separate_XY}
    \begin{split}
        & X(t) =  v_0t + a [n_{r\ell}(t)+n_{m\ell}(t)-n_{\ell m}(t)-n_{\ell r}(t)],\\&
        Y(t) = Y_0 + u_0 t + a [n_{rm}(t)+n_{r\ell}(t)-n_{mr}(t)-n_{\ell r}(t)].
    \end{split}
\end{align}
Let us define the fluctuations around the mean displacements as $\boldsymbol{\Delta} X(t)=X(t)-\langle X(t) \rangle$ and  $\boldsymbol{\Delta} Y(t)=Y(t)-\langle Y(t) \rangle$. To obtain the correlations in the fluctuations of the two particles, it is easier to compute the fluctuation $\langle (Y(t)-X(t))^2 \rangle_c=\langle (\boldsymbol{\Delta} Y(t)- \boldsymbol{\Delta} X(t))^2 \rangle$. We note that 
\begin{align}
  \boldsymbol{\Delta} Y(t)- \boldsymbol{\Delta} X(t)=a[\boldsymbol{\Delta} n_{r m}(t)+\boldsymbol{\Delta} n_{\ell m}(t)-\boldsymbol{\Delta} n_{m r}(t)-\boldsymbol{\Delta} n_{m \ell} (t)],  
\label{delxy}
\end{align}
where $\boldsymbol{\Delta} n_{\ell m}(t)=n_{\ell m}(t)-\langle n_{\ell m} (t)\rangle$ and others are defined accordingly. In the limit where $N\to \infty$ and $\bar{N}$ is finite, the first two terms will dominate at large times, since the last two terms are bounded by the total number of particles in the region $(0,Y_0)$. It is easy to compute $\langle n_{\ell m}(t) \rangle$, noting that it is just the difference of the net flux of particles across the line segments $\{v_0s, 0\leq s \leq t\}$ and $\{y_0+u_0s,0\leq s \leq t\} $. We discuss first the case of  homogeneous initial condition, $\bar{\varphi}(x)=\varphi_0$, for which the expression is simply given by  
\begin{align}
\langle  n_{\ell m}(t) \rangle&=\varphi_0t\int_{v_0}^\infty dw (w-v_0)\mc{h}(w)-\varphi_0 t \int_{u_0+y_0/t}^\infty dw [w- (u_0+y_0/t)] \mc{h}(w), \\
&=\varphi_0t\left[\bar{\msc{F}}(v_0)-\bar{\msc{F}}(u_0+y_0/t)\right],
\label{<n_elm>}
\end{align}
where we recall $y_0=Y_0-(\bar{N}+1)a$.
Similarly,
\begin{align}
\langle  n_{r m}(t) \rangle&=\varphi_0 t\int_{-\infty}^{u_0} dw (u_0-w)\mc{h}(w)-\varphi_0 t \int_{-\infty}^{v_0-y_0/t} dw [w- (u_0+y_0/t)] \mc{h}(w), \\
&=\varphi_0 t\left[ \msc{F}(u_0)-{\msc{F}}(v_0-y_0/t)\right],
\label{<n_rm>}
\end{align}
where ${\msc{F}}(v)$ and $\bar{\msc{F}}(v)$ are defined as
\begin{align}\label{defining F Fbar of v}
    {\msc{F}}(v)=\int_{-\infty}^{v} dw \mc{h}(w) (v-w), ~~ \text{and} ~~ \bar{\msc{F}}(v) = \int_{v}^{\infty} dw  \mc{h}(w)(w-v).
\end{align}
Since $n_{\ell m}(t)$ and $n_{r m}(t)$ are Poisson processes, considering only the first two terms on the rhs of  Eq.~\eqref{delxy}, at large times we obtain 
\begin{align}
\langle (Y(t)-X(t))^2 \rangle_c & = 
a^2 \varphi_0 [ \bar{\msc{F}}(v_0)-\bar{\msc{F}}(u_0) + \msc{F}(u_0)-\msc{F}(v_0) ]t, \\&
=a^2 \varphi_0 (u_0-v_0)t,
\label{var2p}
\end{align}
where the last step follows after straightforward algebra, and the final result has a simple interpretation. At large times, we can ignore the initial separation $Y_0$ between the two quasiparticles, and the fluctuation in the separation mainly gets its contribution from the net number of particles that have entered the region between the two quasiparticles during time $t$. The fluctuation of this net number is the same as the mean number of particles, which is given by $\varphi_0 (u_0-v_0)t$. We note in particular that for $u_0=v_0$, the variance of the separation does not grow with time $t$, indicating that the two quasiparticles move as a rigid object, as noted already in ~\cite{ferrari2023macroscopic}. However, expanding Eqs.~\eqref{<n_elm>} and \eqref{<n_rm>} to the next order in $y_0/t$, we get 
\begin{align}
\lim_{t\to \infty}\langle (Y(t)-X(t))^2 \rangle_c = a^2 \varphi_0 y_0, \label{var-sep-t-inf}
\end{align}
which is simply the equilibrium fluctuation of the number of rods within a region of size $Y_0$. 

We now present a more detailed calculation that allows us to study the early time behavior of the variance and the dependence on $Y_0$ for inhomogeneous initial conditions. 
 As shown in Appendix \ref{app:2-Trc-calc}, for this case also, one can perform a microscopic calculation as in Sec.~\ref{sec:2} in terms of the crossing probabilities such as $p_{\ell r}, p_{r \ell}$, etc. For two quasiparticles, we need four extra crossing probabilities $p_{\ell m}, p_{m \ell}, p_{rm}$ and $p_{mr}$ where $p_{\ell m}/N$ ($p_{rm}/N$) denote the probability of a point quasiparticle to start from the left (right) of $X_0$ ($Y_0$)  and reach the middle region between $X_0$ and $Y_0$ at time $t$. Similarly, $p_{m \ell}$ and $p_{m r}$ are defined. From Eq.~\eqref{MGF for 2T separated} we find that, in the limit $N \to \infty$, the mean  positions of the two quasiparticles are given by 
\begin{align}\label{eq:2quasiparticle_separate_mean_gn}
    \begin{split}
         \langle X(t) \rangle &= v_0t+a(p_{r\ell}+p_{m\ell}-p_{\ell r}-p_{\ell m}), \cr 
        \langle Y (t)\rangle &= Y_0 + u_0t +a (p_{r\ell}+p_{rm}-p_{\ell r}-p_{mr}),
    \end{split}
\end{align}
and the variances are given by 
\begin{align}\label{eq:2quasiparticle_separate_var_gn}
    \begin{split}
         \langle X^2(t)\rangle_c&=a^2 \big( p_{r\ell} + p_{\ell m} + p_{\ell r} +p_{m\ell}\big)-\frac{1}{\bar{N}}a^2p_{m\ell}^2, \\
         \langle Y^2(t)\rangle_c&=a^2\big( p_{r\ell } + p_{rm} + p_{\ell r} +p_{mr}\big)-\frac{1}{\bar{N}}a^2p_{mr}^2.
    \end{split}
\end{align}
The covariance between the positions of the two quasiparticles turns out to be 
\begin{equation}\label{eq:2quasiparticle_separate_cov_gn}
   \langle X(t)Y(t) \rangle_c = a^2(p_{r\ell}+p_{\ell r})+\frac{1}{\bar{N}}a^2p_{m\ell} ~p_{mr}.
\end{equation}
Hence, the variance of the separation between the two quasiparticles is 
\begin{align}
        \langle (Y(t) -X(t))^2 \rangle_c = a^2(p_{\ell m}+p_{rm}+p_{m\ell}+p_{mr})-\frac{1}{\bar{N}}a^2(p_{m\ell}+p_{mr})^2. \label{var-sep-gen}
\end{align}

\begin{figure}[t]
\centering
\includegraphics[width=7.2cm, height=5.6cm]{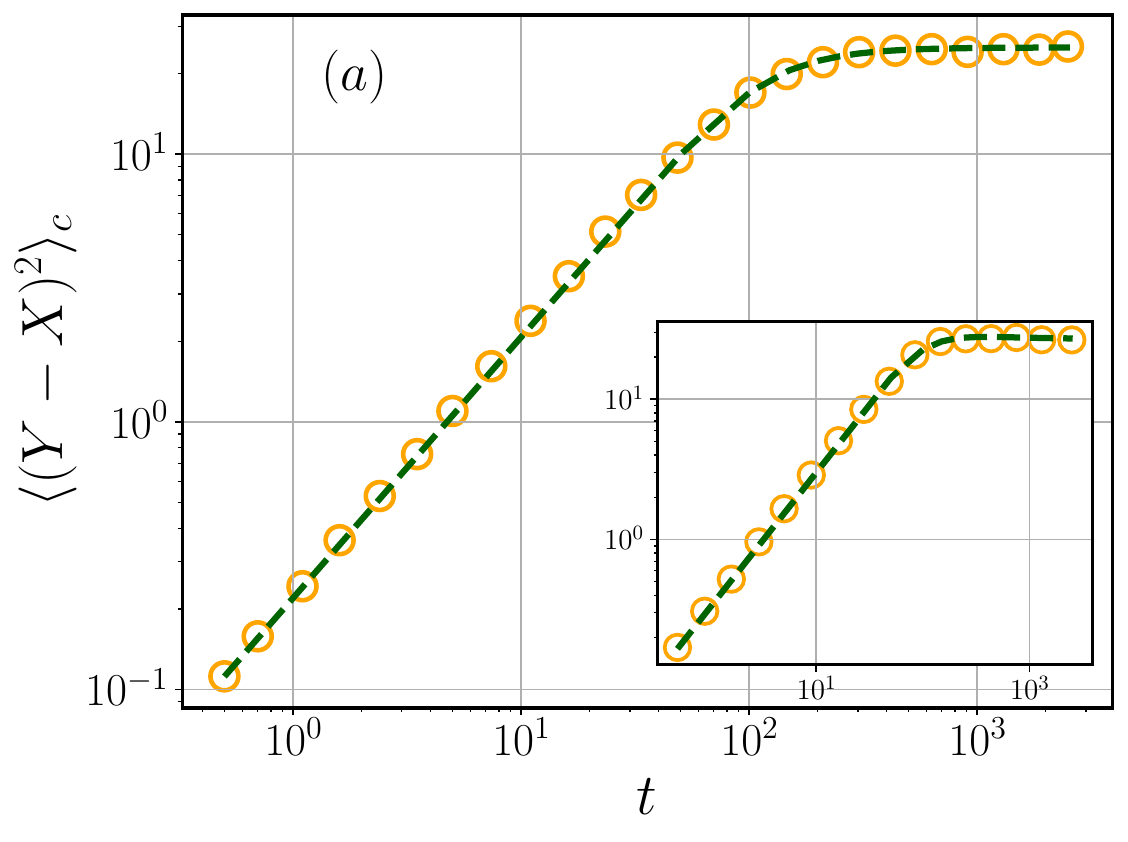}
\includegraphics[width=7.3cm, height=5.7cm]{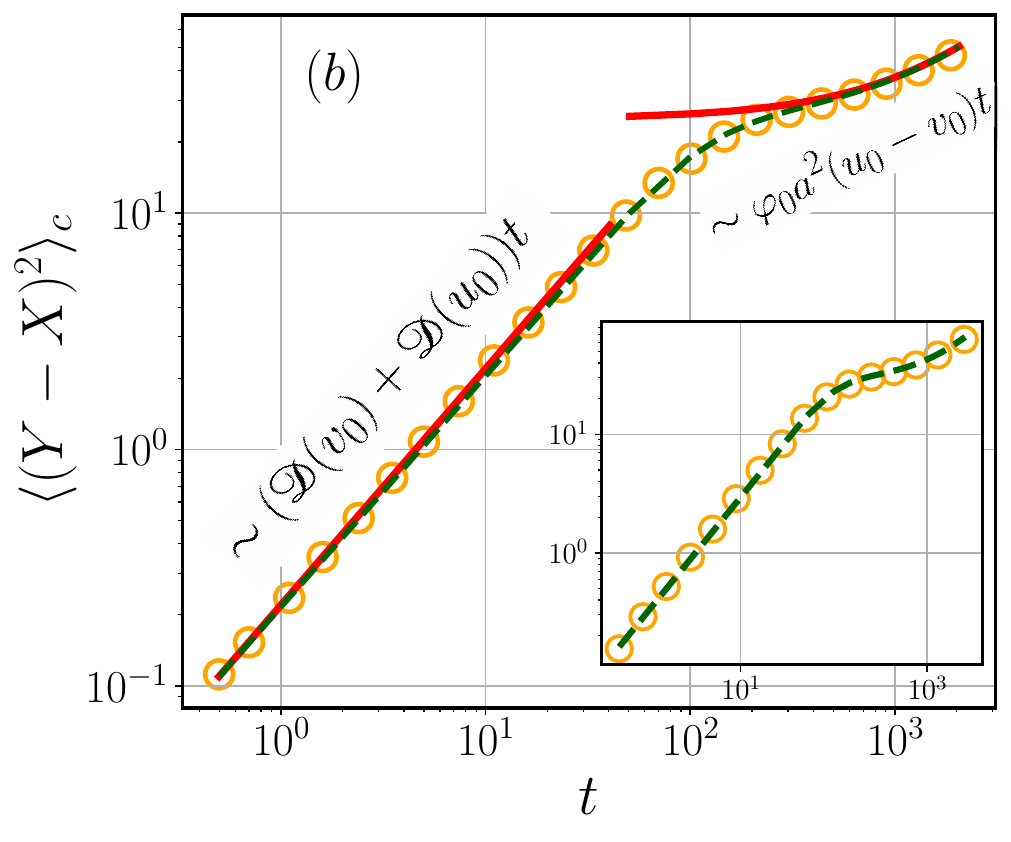}
\vspace*{-.2 cm}
\caption{Numerical simulation results (circles) for the variance of quasiparticle separation are compared with theoretical predictions given in Eq.~\eqref{var-sep-gen}  (dashed lines) for two cases: (a) quasiparticles moving with identical velocities, $v_0=u_0=0.5$;
(b) quasiparticles moving with different velocities $v_0=0.4$, $u_0=0.5$. The main plots correspond to the homogeneous density profile with $\varrho_0=0.4$, while the plots in the inset are for the domain-wall profile with $\varrho_l= 0.4, \varrho_m=0.6, \varrho_r=0.5$. Simulations were performed with $N=5000, a=0.5, T=1$, $Y_0=(\frac{N}{10}+1)a$, with $ \bar{N}=\varrho_0 Y_0$ for the homogeneous case and $\bar{N}=\varrho_m Y_0$ for the domain-wall case. The average has been done over $ 10^4$ independent initial configurations. The red solid lines in Figure (b) here represent the small and large $t$ asymptotics given in Eq. \eqref{t_asymptotic}.}
\label{Fig:AIC_Mean-Var-covar_sep}
\end{figure}

\noindent

Till this point, the results are valid for any general initial mass density profiles $\varphi_{\ell}(x),~ \varphi_r(x), ~ \varphi_m(x)$. As before, the expressions of the mean, variance, and covariance derived above become simple in the special case of a homogeneous initial state in which the mass density profile is uniform throughout the system {\it i.e.,} $\varrho_0$ such that $\varphi_{\ell}=\varphi_r=\varphi_m=\varphi_0=\frac{\varrho_0}{1-a\varrho_0}$. The mean displacements, $\langle X(t) \rangle$ and $\langle Y (t)\rangle$, are given by expressions of the form in Eq.~\eqref{mean-var-Z-homo-a} while the variances are given by:
\begin{align}
    \begin{split}
        & \langle X^2(t)\rangle_c= t {\mc{D}}(v_0)-\frac{a^2 \varphi_0 t^2}{y_0}\Big[{\msc{F}}(v_0)-{\msc{F}}(v_0-y_0/t)\Big]^2, \\&
        \langle Y^2(t)\rangle_c=  t {\mc{D}}(u_0)-\frac{a^2 \varphi_0 t^2}{y_0}\Big[{\bar{\msc{F}}}(u_0)-\bar{{\msc{F}}}(u_0+y_0/t)\Big]^2,
    \end{split}
    \label{exact-var-vo-u0}
\end{align}
\begin{align}
\langle (Y(t)-X(t))^2\rangle_c &= t \Big[{\mc{D}}(v_0) +{\mc{D}}(u_0)-2 a^2\varphi_0\left( {\msc{F}}(v_0-y_0/t)+ \bar{{\msc{F}}}(u_0+y_0/t)\right)  \Big] \cr 
& - \frac{a^2 \varphi_0 t^2}{y_0} \Big[ \Big( {\msc{F}}(v_0)-{\msc{F}}(v_0-y_0/t) \Big) + \Big( \bar{{\msc{F}}}(u_0)-\bar{{\msc{F}}}(u_0+y_0/t) \Big) \Big]^2,~~~~\label{<(Y-X)^2>_c-homo} %
\end{align} 
where, $\mc{u}=\int_{-\infty}^\infty dw ~w \mc{h}({w})$, $y_0=\bar{N}/\varphi_0$. The functions $\msc{F}(v)$ and $\bar{\msc{F}}(v)$ are defined in Eq.~\eqref{defining F Fbar of v} and $\mc{D}(v)$ is defined in Eq.~\eqref{mean-var-Z-homo-b}. The asymptotic behavior of the variance of separation at large and small $t$ is given by 
\begin{align}
\langle (Y(t)-X(t))^2\rangle_c = 
\begin{cases}
\varphi_0 a^2t(u_0-v_0) + a^2\varphi_0 y_0\left[1-\big(\int_{v_0}^{u_0}dw~\mc{h}(w)\big)^2\right]+O\left(\frac{1}{t}\right)\\
~~~~~~~~~~~~~~~~~~~~~~~~~~~~~~~~~~~~~~~~~~~~~~~~~~~~~~~~~~~~~~~~~\text{for~large}~t, \\
t\Big({\mc{D}}(v_0) +{\mc{D}}(u_0)\Big) -\frac{a^2 \varphi_0 t^2}{y_0} \Big( {\msc{F}}(v_0)+\bar{{\msc{F}}}(u_0)\Big)^2+O\left(t\bar{\msc{F}}'\Big( \frac{y_0}{t}\Big)\right)\\
~~~~~~~~~~~~~~~~~~~~~~~~~~~~~~~~~~~~~~~~~~~~~~~~~~~~~~~~~~~~~~~~~\text{for~small}~t,
\end{cases}
\label{t_asymptotic}
\end{align}
where  $\bar{\msc{F}}'(v)=\frac{d \bar{\msc{F}}}{dv}$.

When the two quasiparticles start next to each other {\it i.e.,} $y_0 =0$, or at long times, these correctly reduce to the results derived in Ref.~\cite{ferrari2023macroscopic}. 
The variance of their separation in Eq.~\eqref{<(Y-X)^2>_c-homo} is verified numerically in Fig. \ref{Fig:AIC_Mean-Var-covar_sep} for both homogeneous and domain-wall initial conditions (inset). Figs.~\ref{Fig:AIC_Mean-Var-covar_sep}(a) and (b) correspond to the cases of equal and unequal bare velocities of the quasipartiles. The results for domain-wall initial conditions are discussed in Appendix \ref{mvc-dw}. The excellent agreement between the theory and simulation data provides additional verification of our results. 

\begin{figure}[h]
\centering
\includegraphics[width=10cm, height=7.2cm]{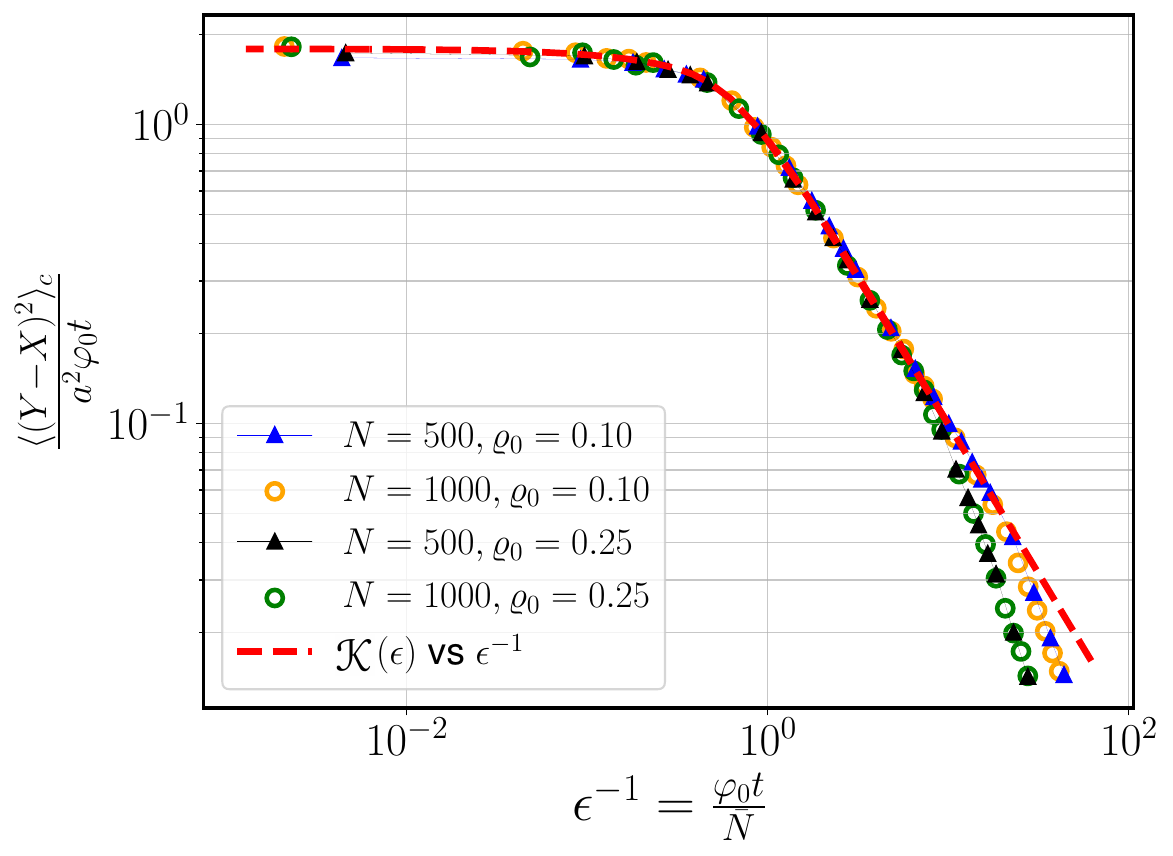}
\vspace*{-.2 cm}
\caption{Scaling collapse of the variance of separation $\frac{\langle (Y-X)^2\rangle_c}{t}$ plotted as a function of time $\epsilon^{-1}=\varphi_0t/ \bar{N}$ in the log scale for different values of $\bar{N}=\varrho_0 ({N}/{2}+1)a$ under homogeneous initial condition. For each $\varrho_0$, the data for different $\bar{N}$ collapses to a scaling curve. The scaling curve for smaller $\varrho_0$ appears to converge to the theoretical scaling function in Eq. \eqref{var-sep-scaling-function} (red dashed line), which corresponds to the limit $\bar{N}/N \to 0$. The parameters used are: $a=0.5$, $v_0=u_0=0.5$, $Y_0=(N/2+1)a$. Average has been done over $10^4$ independent initial configurations for each $N$.} 
\label{Fig:Scaling_variance_d}
\end{figure}

We now focus on the case where the quasiparticles start with separation $Y_0$ and with the same velocity, i.e., $u_0=v_0$.   
The mean separation in this case remains $\langle Y(t)-X(t) \rangle = Y_0 $,
where, recall, $Y_0$ is the initial separation between the two quasiparticles.  On the other hand, the variance of their separation grows linearly with time at small times and saturates at large times (noted previously in Eq.~\eqref{var2p} and Eq.~\eqref{var-sep-t-inf}). The finite-time behavior of the variance is given by the following scaling form 
\begin{subequations}
\begin{align}
\frac{\langle (Y(t)-X(t))^2\rangle_c}{t} = a^2 \varphi_0 ~\msc{K}\left(\frac{\bar{N}}{t\varphi_0}\right), \label{var_sep-scaling}
\end{align}
where, recall $y_0=\frac{\bar{N}}{\varphi_0}$, and 
\begin{align}
\msc{K}(\epsilon)=&2[\msc{F}(v_0)+\bar{\msc{F}}(v_0)-\msc{F}(v_0-\epsilon)-\bar{\msc{F}}(v_0+\epsilon)] 
 \cr & ~~
 -\frac{1}{\epsilon}[\msc{F}(v_0)+\bar{\msc{F}}(v_0)-\msc{F}(v_0-\epsilon)-\bar{\msc{F}}(v_0+\epsilon)]^2.
\end{align}
\label{var-sep-scaling-function}
\end{subequations}
\noindent
In Fig.~\ref{Fig:Scaling_variance_d}, we verify this scaling numerically, where we see that the scaling curve is flat at small $t$ and approaches zero at large $t$, consistent with the linear growth of $\langle (Y(t)-X(t))^2\rangle_c$ at small times and saturation at large times [see Eq.~\eqref{var-sep-t-inf}]. 
At small times, the two quasiparticles move independently of each other and individually contribute to the linear growth of the separation. With increasing time, their motion starts to become correlated because they start crossing the same background quasiparticles. 

The scaling collapse suggests that the departure from linear growth occurs on a time scale $O(y_0)$, which can be understood as follows. Recall that $y_0={\bar{N}}/{\varphi_0}$, where $\bar{N}$ is the initial number of rods between the two quasiparticles. A background quasiparticle that has just crossed $X(t)$ must interact with the quasiparticles located between $X(t)$ and $Y(t)$ before colliding with $Y(t)$. Since the separation between $X(t)$ and $Y(t)$ is $O(y_0)$, the typical time required for a background quasiparticle to traverse this region and establish the first correlation event is also of this order.
Following Ref.~\cite{ferrari2023macroscopic}, one can argue that saturation of the variance to the value $a^2\varphi_0 y_0$ occurs on a time scale $O(y_0^2)$. The quasiparticles have an initial separation $Y_0=y_0(1+a\varphi_0)$ and move with the same bare velocity $v_0$. They need to fluctuate by $O(y_0)$  to become fully correlated, and this happens on a time scale of $O(y_0^2)$. Over this time period, each of them suffers $O(y_0^2)$ number of collisions,  most of which are with the same background quasiparticles, except for an $O(y_0)$ collisions~\cite{ferrari2023macroscopic}. 
 Hence, on this time scale, the two quasiparticles essentially suffer the same displacements and move collectively as a rigid body, while their center of mass executes a Brownian motion.

\section{$\langle X^2\rangle_c$ and $\langle (Y-X)^2 \rangle_c$  from Euler GHD in homogeneous background}
\label{corr-frm-EGHD}
It has been argued that the correlation among the fluctuations of the phase space densities at two different phase space points, charaterized by the correlations in the noise $\zeta(X,v,t)$ essentially originates from the fluctuations of $\mc{f}(X,v,0)$ in the initial state that gets carried to time $t$ deterministically by Euler equations \cite{doyon2023emergence, doyon2023ballistic, kundu2025macroscopic, hubner2024diffusive,hubner2025hydrodynamics,bulchandani2024revised,doyon2025bhydrodynamic,PhysRevB.109.024417,PhysRevLett.128.160601,PhysRevE.111.024141,equations}. To demonstrate this fact, in this section, we show how the initial correlation determines the fluctuations in the displacement of a quasiparticle at time $t$.

\noindent
We start by defining an empirical density in the single particle phase space
\begin{align}
    \mc{f}(X,v,t)=\sum_{i=1}^{\msc{N}} \delta(X-X_i(t))~\delta(v-v_i). \label{f_emprcl}
\end{align}
In the ballistic space-time scale the PSD $\mc{f}(X,v,t)$ evolves according to Euler GHD equation \cite{hubner2025hydrodynamics,singh2024thermalization,Mrinal_2024_HR,hubner2025diffusive,Dobrushin1983,Dobrushin1990}
\begin{align}
\partial_t\mc{f}(X,v,t) +\partial_X(v_{\rm eff}(v)\mc{f}(X,v,t))=0,~~\text{with}~~v_{\rm eff}(v) = \frac{v-a\int du~u \mc{f}(X,u,t)}{1-a\int du~\mc{f}(X,u,t)}.
\label{f-rod-eu}
\end{align}
Using the following transformations, 
\begin{align}
f(x(X),v,t)&=\frac{\mc{f}(X,v,t)}{1-a\varrho(X,t)}, \label{def:fr(X,y,t)}\\
x(X)&=X-a\int_{-\infty}^XdZ~\varrho(Z,t), \label{def:X-x-hd}
\end{align}
with $\varrho(Y,t)=\int dv~\mc{f}(Y,v,t)$,  Eq.~\eqref{f-rod-eu} becomes 
\begin{align}
\partial_t f(x,v,t)+v\partial_xf(x,v,t)=0, \label{Eu-eq-f-pp}
\end{align}
which is the Euler equation for the phase space density of point particles \cite{singh2024thermalization}. Note that the transformation in the above equation is essentially the continuous limit of the mapping in Eq.~\eqref{eq:map-hr-hp}.  The advantage of going to the point particle representation is that the Euler equation can now be solved just by boosting the initial PSD $f(x,v,0)$:
\begin{align}
f(x,v,t)=f(x-vt,v,0). \label{sol:Eu-f}
\end{align}
The position of the quasiparticle can easily be expressed in terms of the mass density of point particles as 
\begin{align}
X(t)=X(0)+v_0t+a\left[ \int _{-\infty}^{x(0)+v_0t}dy~\varphi(y,t) - \int_{-\infty}^{x(0)}dy~\varphi(y,0)\right], \label{def:X(t)-hd-p}
\end{align}
where $\varphi(x,t)=\int dv f(x,v,t)$ represents the mass density of the hard-point gas and is related to $\varrho(X,t)$ as $\varphi(x(X),t)=\frac{\varrho(X,t)}{1-a\varrho(X,t)}$ and  $X(0)=x(0) + a \int_{0}^{x(0)} dy~ \varphi(y,0)$ is the position of the hard rod corresponding to the point particle at  position $x(0)$.  Note that the first term inside the square brackets represents the number of points of particles below the location $x(t)=x(0)+v_0t$ at time $t$, which is essentially the same as the number of rods below the position $X(t)$ at time $t$. The same interpretation holds for the second term inside the square brackets. Together, these two terms then provide the net number of rods that have crossed the quasiparticle from right to left during time $t$.  

Let us fix the initial condition of the point particle as $x(0)=0$, this  means that the initial position of the hard rod $X(0)$ will be a  fluctuating variable. In this case we can  compute the mean and variance of the quasiparticle displacement $X(t)-X(0)$ and note that for homogeneous initial condition these are same as the mean and variance of quasiparticle position with its initial position fixed, as was considered in Sec.~\ref{sec:3}. However, the initial conditions are  slightly different from the ones discussed in the earlier sections, where $X(0)=0$ was fixed and an equal number of particles in equilibrium on the two sides. However, we expect that large time results to be unchanged, as indeed will be confirmed. Note that $f(x,v,t)$ is the phase space density of the point particles that evolved from the initial density $f(x,v,0)$. Since the initial profile is randomly chosen from a homogeneous equilibrium state, the time-evolved profile $f(x,v,t)$ is also random, causing the quasiparticle displacement,  $\Delta X(t) = X(t)-X(0)$, to fluctuate through Eq.~\eqref{def:X(t)-hd-p}. Using  $\bar{f}(y,v,t)=\langle f(y,v,t)\rangle =\varphi_0 \mc{h}(v)$, it is easy to see that $\langle X(t)-X(0) \rangle=v_{\rm eff}t$ as in Eq.~\eqref{mean-var-Z-homo}. The variance of the displacement of the quasiparticle at time $t$ is related to the unequal space-time correlation of the PSD,  $f(x,v,t)$.

More precisely, the variance of $\Delta X(t) $ can now be written as integrals over the space-time correlation of the point particle  densities:
\begin{align}
\langle [\Delta X(t) ]^2\rangle_c =\langle \mc{A}(0)^2 \rangle + \langle \mc{A}(t)^2 \rangle -2 \langle \mc{A}(t)\mc{A}(0) \rangle, \label{sigma_x^2(t)-hd}
\end{align}
where $\mc{A}(t)=\int _{-\infty}^{v_0t}dy~\delta \varphi(y,t)$ with $\delta \varphi(y,t)= \varphi(y,t)-\varphi_0$ and 
\begin{align}
\langle \mc{A}(t)\mc{A}(t')\rangle&= a^2 \int_{-\infty}^{v_0t}dy\int_{-\infty}^{v_0t'}dy'~\langle \delta \varphi(y,t) \delta \varphi(y',t') \rangle. \label{<AA>}
\end{align}
We assume the hard rod gas starts in a homogeneous state with mass density $\varrho_0$ and velocities chosen from a symmetric distribution $\mc{h}(v)=\mc{h}(-v)$. For this case, it is easy to prove that the initial correlation for the phase space density of point particles is  \cite{kundu2025macroscopic,kethepalli2025ballistic,kundu2020dynamical}
\begin{align}
\langle f(x,v,0)f(y,u,0)\rangle = \varphi_0~\delta(x-y)\delta(v-u)~\mc{h}(v). \label{<ff>(0)}
\end{align}
As the particles move, this correlation also evolves. Since we are interested in the correlation at the Euler space-time scale (space is proportional to time), it should just be the one obtained by propagating the initial random PSD profile $f(x,v,0)$  ballistically by the Euler equations {\it i.e.} $f(x,v,t) =f(x-vt,v,0)$. By doing so, we discard the fluctuations over a small space-time scale, possibly arising due to coarse-graining. Hence, we write
\begin{align}
\begin{split}
\langle \delta \varphi(y,t) \delta \varphi(y',t') \rangle&= \int dv \int du~\langle f(y,v,t)f(y',u,t')\rangle_c, \cr 
&= \int dv \int du~\langle f(y-vt,v,0)f(y'-ut',u,0)\rangle_c. \label{<vphi(t)vphi(t')>}
\end{split}
\end{align}
Now, inserting the initial correlation in the above equation and performing the integrals over the velocities, we get 
\begin{align}
\langle \delta \varphi(y,t) \delta \varphi(y',t') \rangle = \varphi_0 \frac{1}{|t -t'|}~\mc{h}\left(\frac{y-y'}{|t-t'|}\right),
\label{<phi(t)phi(t')>_eq}
\end{align}
which also gives $\langle \delta \varphi(y,t) \delta \varphi(y',t) \rangle= \varphi_0 \delta(y-y')$. In terms of the mass density of hard rods, this correlation reads \cite{kundu2025macroscopic,doyon2017dynamics}
\begin{align}
  \langle \varrho(Y,t)\varrho(Y',t')\rangle_{c} = \varrho_0(1-a\varrho_0)^3 \frac{1}{|t -t'|}~\mc{h}\left((1-a\varrho_0)\frac{Y-Y'}{|t-t'|}\right).
\end{align}
Using the expression of the correlations from Eq.~\eqref{<phi(t)phi(t')>_eq} in  Eq.~\eqref{<AA>},  one can show that 
\begin{align}
2\langle \mc{A}(t)\mc{A}(0) \rangle&= \left[\langle \mc{A}(0)^2 \rangle+\langle \mc{A}(t)^2 \rangle \right] \cr 
&-\left\{ \int_{v_0t}^\infty dy\int_{-\infty}^\infty dy' +\int_{-\infty}^{v_0t} dy\int_{-\infty}^\infty dy' \right\} \frac{1}{t}\mc{h}\left( \frac{y-y'}{t}\right).
\end{align}
Further, using this relation in Eq.~\eqref{sigma_x^2(t)-hd}, and simplifying, one gets the same expression as in Eq.~\eqref{mean-var-Z-homo-a} obtained previously using the microscopic approach.
Following a similar procedure, one can also compute two-time auto correlation $\langle \Delta X(t_1) \Delta X(t_2)\rangle_c$  and reproduce the result in Eq.~\eqref{eq:2time_corr_1quasiparticle_expression}.

The HD procedure, described above, can also be extended to compute the covariance between two quasiparticles $\langle \Delta X(t) \Delta Y(t) \rangle_c$. As before, we start with two point particles, one at position $x(0)=0$ and another at $y(0)=y_0$, with velocities $v_0$ and $u_0$, respectively.  The corresponding quasiparticles are located at $X(0)$ and $Y(0)$.
The position $X(t)$ of the first quasiparticle at time $t$ is defined in Eq.~\eqref{def:X(t)-hd-p}. In the same way, one can define the position $Y(t) = Y(0) + \Delta Y(t)$ of the second quasiparticle where 
\begin{align}
\Delta Y(t)=u_0t + a\left[ \int_{-\infty}^{y_0+u_0t}dy~\varphi(y,t) - \int_{-\infty}^{y_0}dy~\varphi(y,0)\right]. \label{def:Y(t)-hd-p}
\end{align}
The covariance $\langle \Delta X(t) \Delta Y(t) \rangle_c$ can now be written as
\begin{align}
\begin{split}
\langle \Delta X(t) \Delta Y(t) \rangle_c=& \mc{E}(y_0+u_0t,t~;~v_0t,0) - \mc{E}(y_0+u_0t,t~;~0,0) \cr 
&~- \mc{E}(y_0,0~;~v_0t,t) +\mc{E}(y_0,0~;~0,0),
\end{split}
\label{def:S_XY^2-hd}
\end{align}
where
\begin{align}
\mc{E}(z,t;z',t')= a^2 \int_{-\infty}^zdy \int_{-\infty}^{z'} dy' \langle \delta \varphi(y,t) \delta \varphi(y',t') \rangle.
\end{align}
Inserting the form of the correlation from Eq.~\eqref{<phi(t)phi(t')>_eq} in Eq.~\eqref{def:S_XY^2-hd} and performing the integrals, one obtains $\langle \Delta X(t) \Delta Y(t)\rangle_c$ explicitly. Using this correlation  along with Eq.~\eqref{exact-var-vo-u0} one can compute $\langle(\Delta Y(t)-\Delta X(t))^2\rangle_c$ which exactly reproduces the 
result in Eq.~\eqref{<(Y-X)^2>_c-homo}.

\section{Phenomenological derivation of fluctuating hydrodynamics}\label{sec:quasiparticles_fluc_hd}

The results of the previous two sections \ref{sec:3} and \ref{sec:two_quasiparticles} indicate that quasiparticles in a homogeneous gas of hard rods effectively move as Brownian particles with an effective drift but are driven by correlated noises. The effective equations of motion of such particles are 
\begin{align}
    \frac{dX_i}{dt}= \bar{v}_i +\xi_{i}(t),~~\text{for}~~i=1,2,...,\msc{N},~~\text{with}~~\bar{v}_i=\frac{v_i}{1-a\varrho_0}, \label{Brownian-QP}
\end{align}
where $\varrho_0$ is the mass density of the hard rod gas and $v_i$ are the individual bare velocities of the particles.  The stochastic variables  $\xi_i(t)$ are mean-zero white Gaussian noises with correlations 
\begin{align}\label{eq:noise_corr}
\begin{split}
    \langle \xi_i(t)\xi_j(t')\rangle &= \frac{\mc{G}(v_i,v_j) }{\sqrt{\mc{D}(v_i)}\sqrt{\mc{D}(v_j)}}~\delta(t-t'),\cr
    \langle \xi_i(t)\xi_i(t')\rangle &=\mc{D}(v_i) ~\delta(t-t'),
    \end{split}
    ~~\text{for}~~i,j= 1,2,...,\msc{N},
\end{align}
where $\mc{D}(v)$ is given in Eq.~\eqref{mean-var-Z-homo-b} and $\mc{G}(v,u)$ is  defined by
\begin{align}\label{eq:G_uv}
\mc{G}(u,v) =\frac{1}{2}\left( \mc{D}(v)+\mc{D}(u)- a^2 \varphi_0|v-u|\right).
\end{align}
By considering the hard rod gas as a collection of such Brownian particles,  in this section,  we provide a phenomenological derivation of a fluctuating hydrodynamic equation derived rigorously in \cite{ferrari2023macroscopic}. We follow the Dean-Kawasaki approach~\cite{dean1996langevin,illien2024dean} for deriving the fluctuating hydrodynamic equations.

We are interested in finding the stochastic differential equation satisfied by the fluctuating PSD $\mc{f}(X,v,t)$ (see Eq.~\eqref{f_emprcl}) in the thermodynamic limit. To proceed, it seems convenient to go to the Fourier space. We define
\begin{align}
    \hat{\mc{f}}(k,v,t)=\int_{-\infty}^{\infty}dX ~e^{\mc{i} kX}~\mc{f}(X,v,t)=\sum_{j}e^{\mc{i} kX_j(t)}~\delta(v-v_j).
\end{align}
At time $t+dt,$ we have
\begin{align}
    \hat{\mc{f}}(k,v,t+dt)&= \sum_je^{\mc{i} kX_j(t+dt)}~\delta(v-v_j), \cr 
    &=\sum_je^{\mc{i} k[X_j(t)+\bar{v}_jdt+\delta \xi_j(dt)]}~\delta(v-v_j), \\
  &= \sum_je^{\mc{i} kX_j(t)}\left[ 1 +\mc{i} k ~\bar{v}_jdt + \mc{i} k ~\delta \xi_j(dt) -\frac{k^2}{2}(\delta \xi_j)^2 + O(dt^2) \right]\delta(v-v_j), \notag
\end{align}
where $\delta \xi_j(dt) = \int_t^{t+dt}dt'~\xi_j(t')$.
To leading order in $dt$, we can write $(\delta \xi_j(dt))^2 = \mc{D}(v_j)dt$ and  retaining terms  up to linear order in $dt$ one gets, 
\begin{align}\label{eq:fhd_fourier fkvt}
    \partial_t\hat{\mc{f}}(k,v,t)=\mc{i} k \bar{v}\hat{\mc{f}}(k,v,t)-\frac{k^2}{2}~ {\mc{D}}(v)\hat{\mc{f}}(k,v,t) +\mc{i} k~ \zeta_k(v,t),
\end{align}
with 
\begin{align}
     \zeta_k(v,t)=\frac{1}{dt}\sum_{j}e^{\mc{i} k X_j(t)}~\delta(v-v_j)\delta \xi_j(dt).
\end{align}
Performing inverse Fourier transform, Eq. \eqref{eq:fhd_fourier fkvt} immediately yields
\begin{align}
    \partial_t\mc{f}(X,v,t)&=-\bar{v}~\partial_X\mc{f}(X,v,t)+\frac{{\mc{D}}(v)}{2}~\partial_X^2~ \mc{f}(X,v,t)+\partial_X ~ \zeta(X,v,t), 
    \label{eq:fhd_eq_fxvt1}
\end{align}
where 
\begin{align}
\zeta(X,v,t)= \lim_{dt \to 0} \frac{1}{dt}\sum_j\delta(X-X_j(t))\delta(v-v_j)\delta \xi_j(dt). \label{zeta(X,v,t)}
\end{align}
It is easy to check that $\langle \zeta(X,v,t)\rangle =0$ and 
\begin{align}
\langle \zeta(X,v,t)\zeta(Y,v,t')\rangle &= \mc{D}(v)\mc{f}(X,v,t) \delta(X-Y)  \delta(t-t') , \cr
\langle \zeta(X,v,t)\zeta(Y,u,t)\rangle &=  \mc{G}(u,v)\mc{f}(X,v,t)\mc{f}(Y,u,t) \delta(t-t'),
\end{align}
where we recall that $\varphi_0=\frac{\varrho_0}{1-a\varrho_0}$. It is convenient to define the white noise $\dot{W}_t(v)$ via 
$\zeta(X,v,t)=\mc{f}(X,v,t)\sqrt{{\mc{D}}(v)}~\dot{W}_t(v),$ such that 
\begin{align}
    \langle \dot{W}_t(v) \rangle = 0, ~\langle \dot{W}_t(v)\dot{W}_{t'}(u)\rangle=\frac{\mc{G}(v,u)}{\sqrt{{\mc{D}}(v){\mc{D}}(u)}}~\delta(t-t'). \nonumber
\end{align}
Finally, we obtain the following fluctuating hydrodynamic equation of the PSD $\mc{f}(X,v,t)$  
\begin{align}
\begin{split}
    \partial_t\mc{f}(X,v,t)=&-v_{\rm eff}(v)~\partial_X\mc{f}(X,v,t)+\frac{{\mc{D}}(v)}{2}~\partial_X^2~ \mc{f}(X,v,t)  \\ 
    &~~~~~~~~~~~~~~~~~~+\partial_X~\mc{f}(X,v,t)\sqrt{{\mc{D}}(v)}~\dot{W}_t(v),
    \end{split}
    \label{eq:fhd_eq_fxvt2}
\end{align}
where $v_{\rm eff}(v)=\bar{v}=v/(1-a \varrho_0)$. 
Performing a coordinate transformation to the Euler frame $X \to X-\bar{v}t$, we get an equation that has exactly the form of the fluctuating hydrodynamic equation on the diffusive space-time scale obtained by Ferrari and Olla~\cite{ferrari2023macroscopic}. However, note that our derivation using the Dean-Kawasaki approach did not require any coarse-graining either in space or in time. Hence, it is not clear that they are the true fluctuating hydrodynamic equations that would follow from a coarse-grained theory.  For further discussions on this point, see Sec.~\eqref{sec:conclusion}.

\section{Initial state with quenched positions}\label{sec:two quasiparticles quenched}
In Section \ref{sec:two_quasiparticles}, we have studied the evolution of two quasiparticles in a background of many hard rods under the annealed initial condition. Here, we extend our microscopic analysis of quasiparticle diffusion to the case of quenched initial conditions, where the initial positions of the particles are quenched, meaning they are fixed according to a specific pattern. Once again, we first decide the positions in the point particle picture and then transform to hard rod coordinates using Eq.~\eqref{inv-map-to-hpg-line}. We consider a quenched configuration $\{\bar{x}_i~|~\bar x_i <\bar x_{i+1}; i=1,2\dots \msc{N}-1\}$ for the point particles such that in the thermodynamic limit, they correspond to a well-defined macroscopic mass density profile $\varphi_q(\bar x)$. The velocities of the particles at these positions are assumed to be randomly sampled from a distribution ${\mc{h}}(v).$  Hence, the joint distribution of the positions and momenta of the point  particles can be formally written as 
\begin{align}
    \mathbb{P}_q(\{x_i,v_i\}|0)=\prod_{i=1}^N\delta( x_i-\bar x_i)~\mc{h}(v_i).
    \label{eq:quenched_IC_define}
\end{align}
Here, we compute the mean, variance, and covariance of the positions of two quasiparticles, once again initially placed at $X(0)=X_0=0$ and $Y(0)=Y_0>0$ such that there are $\bar {N} \le (Y_0-a)/a $ rods in between them. 

The computation of these statistical properties of the two quasiparticles for general quenched initial conditions is discussed in detail in Appendix \ref{app:2quasiparticle_quenched_prob}. The expressions derived in Eqs. \eqref{eq:2quasiparticle_separated_quenched_mean} - \eqref{eq:2quasiparticle_separated_quenched_var_covar} remain valid for arbitrary initial densities of the background rods. Here we only present the results for the specific case of a homogeneous initial state with density $\varrho_0$ and $Y_0=a$ ({\it i.e.} the limit corresponds to zero separation between the quasiparticles). This also corresponds to a uniform density for the point particles, with value $\varphi_0=\frac{\varrho_0}{1-a\varrho_0}$. Under this setting, we find that the mean positions are still given by Eq.~\eqref{mean-var-Z-homo-a}. 
The variance and covariance turn out to be 
\begin{figure}[t]
\centering
\includegraphics[width=4.9cm, height=4.5cm]{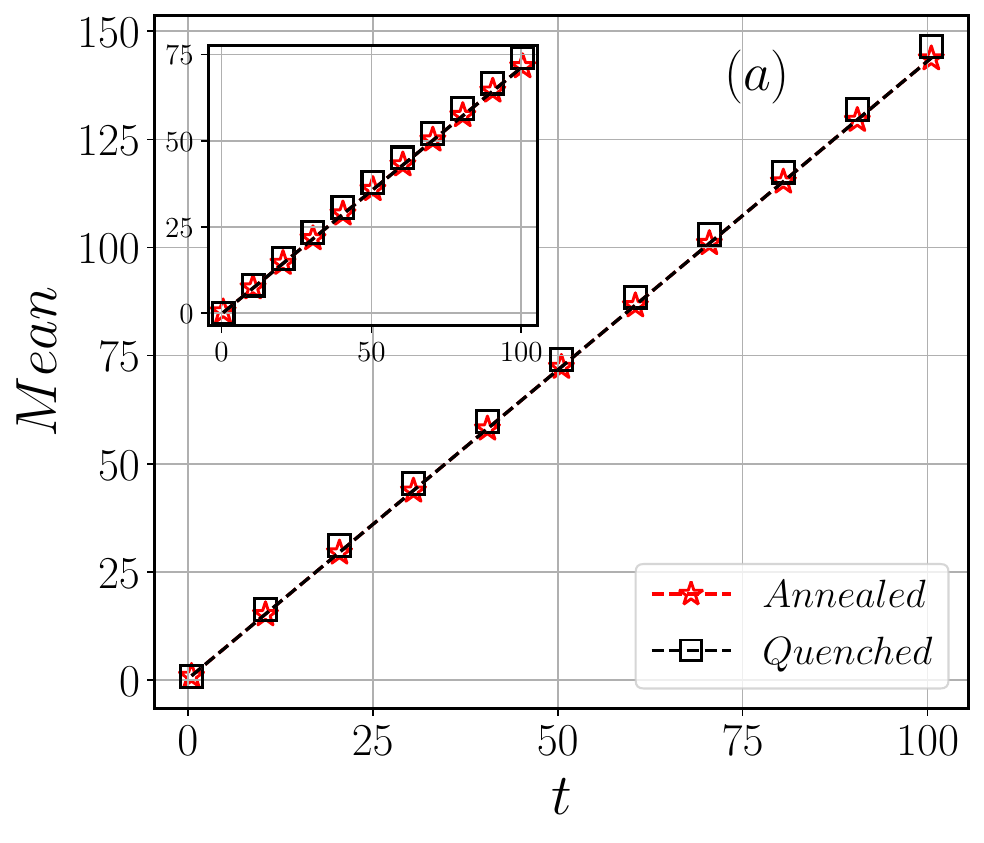}
\includegraphics[width=4.9cm, height=4.5cm]{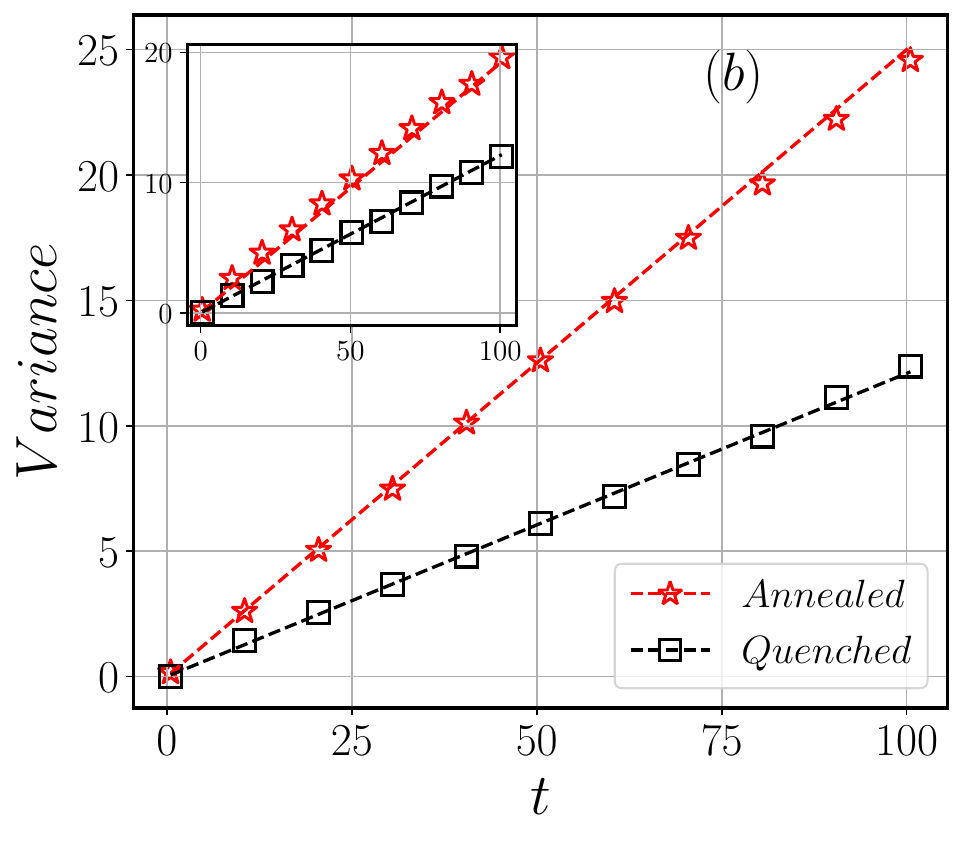}
\includegraphics[width=4.9cm, height=4.5cm]{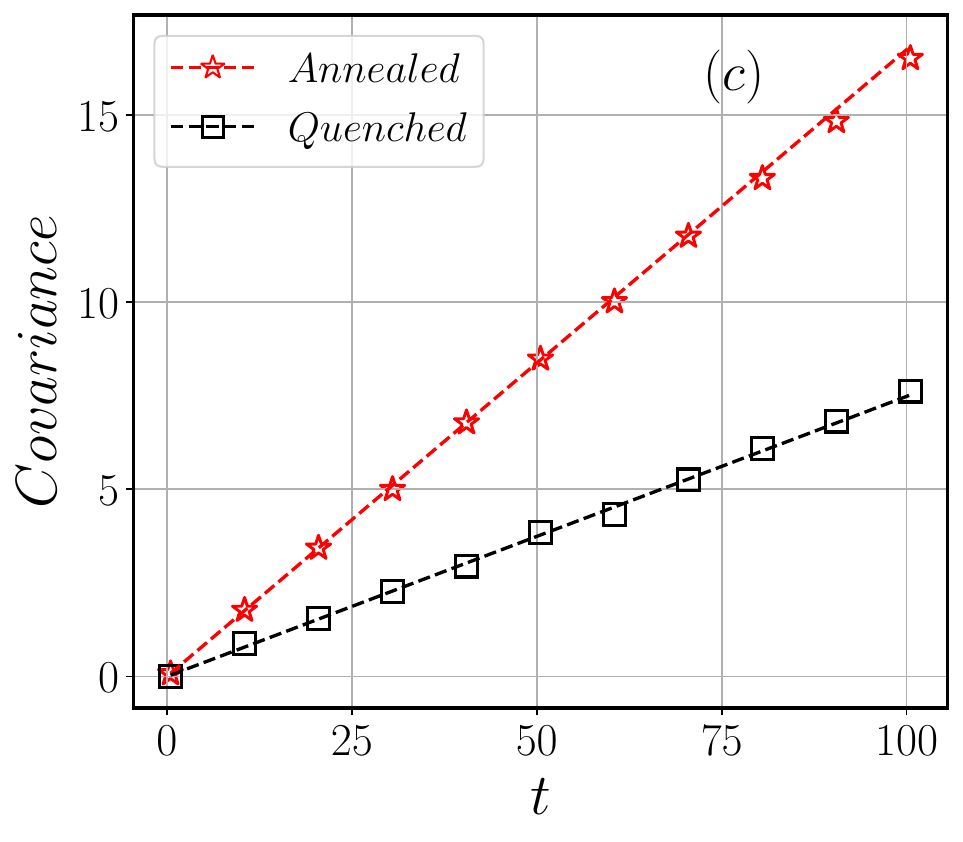}
\caption{Numerical simulation (black squares) and theoretical results (black dashed lines) for (a) mean, (b) variance, and (c) covariance of two quasiparticles $X(t), Y(t)$ from quenched initial condition following Eqs.~\eqref{eq:2quasiparticle_origin_var_QIC_uniform}-\eqref{eq:2quasiparticle_origin_covar_QIC_uniform}. Insets show the corresponding results for quasiparticle $X(t)$. Results from quenched initial conditions are compared with those for annealed initial conditions (red stars: simulations; red dashed lines: theory). The parameters are $ a=0.5, v_0=0.5, ~ u_0=1.0, ~T=1$, $N=3000,\varrho_0 =0.6$ and $Y_0=a$. The average has been done over $10^4$ realizations.}\label{Fig:QIC_Mean-Var-covar_sep}
\end{figure}

\begin{align}\label{eq:2quasiparticle_origin_var_QIC_uniform}
\begin{split}
    \langle X^2(t)\rangle_c=t\mc{D}(v_0) -a^2\frac{\varrho_0t}{1-a\varrho_0}\Bigg[\int_{-\infty}^{v_0}dw~ \msc{H}_+^2(w)+\int_{u_0}^{\infty}dw~\msc{H}_-^2(w-u_0+v_0)\Bigg],\\
    \langle Y^2(t)\rangle_c=t\mc{D}(u_0) -a^2\frac{\varrho_0t}{1-a\varrho_0}\Bigg[\int_{u_0}^{\infty}dw~ \msc{H}_-^2(w)-\int_{-\infty}^{v_0}dw~\msc{H}_+^2(w+u_0-v_0)\Bigg],
    \end{split}
\end{align}
\begin{align}\label{eq:2quasiparticle_origin_covar_QIC_uniform}
    \langle X(t)Y(t) \rangle_c=t\mc{G}(u_0,v_0)-a^2\frac{\varrho_0t}{1-a\varrho_0}\Bigg[&\int_{-\infty}^{v_0}dw~\msc{H}_+(w)\msc{H}_+(w+u_0-v_0)\nonumber \\
    &-\int_{u_0}^{\infty}dw~ \msc{H}_-(w)\msc{H}_-(w-u_0+v_0)\Bigg],
\end{align}
where 
\begin{align}
\msc{H}_\pm(w) = \int_{-\infty}^\infty dw'~\theta\left(\pm(w-w')\right)~\mc{h}(w'),
\end{align}
and recall $\mc{D}(v)$ and $\mc{G}(u,v)$ are the same as defined in Eqs. \eqref{mean-var-Z-homo-b}  and  \eqref{eq:G_uv}. The results in Eqs. \eqref{eq:2quasiparticle_origin_var_QIC_uniform} - \eqref{eq:2quasiparticle_origin_covar_QIC_uniform}  corresponds to the distribution of the quasiparticle positions in
thermodynamic limit. We verify these results numerically in Fig.~\ref{Fig:QIC_Mean-Var-covar_sep} for homogeneous densities of the background rods. For comparison, we plot the mean, variance, and covariance of quasiparticle positions obtained using annealed and quenched initial conditions, both corresponding to the same initial density profiles. Our numerical simulations show that, as expected, the variances and covariances in the quenched case are consistently smaller than those in the annealed case.

\section{Conclusion}\label{sec:conclusion}
We have studied the dynamics of quasiparticles in the background of many hard rods from a microscopic point of view. For a certain type of initial conditions, we compute explicit expressions of the mean, variance, and covariance of the quasiparticles. For a homogeneous background of the hard rod gas, our computation reproduces the results previously derived by Ferrari and Olla \cite{ferrari2023macroscopic}. In addition to providing numerical verification of these results, we extend the computations of these quantities to inhomogeneous cases as well. Our results, as was also claimed in \cite{ferrari2023macroscopic}, indicate that the quasiparticles effectively perform strongly correlated Brownian motions with velocity-dependent diffusion constants. Such correlations make two quasiparticles with the same velocity move effectively as a rigid body. The fluctuations in the phase space densities essentially originate from the initial fluctuations that ballistically evolved to time $t$ via Euler GHD. To demonstrate this fact, we rederive the variance and covariance of the quasiparticle positions using Euler scale correlations in the mass densities for the homogeneous case.  

The correlations between two quasiparticles in the homogeneous case have been used in \cite{ferrari2023macroscopic} to obtain a fluctuating hydrodynamic equation for the PSD of the rods. On the phenomenological level, considering an approximate description of a homogeneous hard rod gas as a gas of non-interacting quasiparticles performing correlated Brownian motions as mentioned above, we have attempted to provide an alternative physical derivation of the fluctuating hydrodynamic equations obtained in \cite{ferrari2023macroscopic}. 
However, we now point to a number of puzzles related to the obtained fluctuating hydrodynamic
equations.
The HD equation for the average single particle distribution $\bar{\mc{f}}(X,v,t) = \langle \mc{f}(X,v,t) \rangle$ of hard rods with Navier-Stokes correction has been derived in several works \cite{boldrighini1997one, doyon2017dynamics, hubner2025diffusive, hubner2025hydrodynamics}. In linear order in deviation from global equilibrium $\bar{\mc{f}}(X,v,t) =\varrho_0 \mc{h}(v) + \tilde{\mc{f}}(X,v,t)$, the HD equation reads 
\begin{align}
    \partial_t &\tilde{\mc{f}}(X,v,t) +\bar{v} \partial_X\tilde{\mc{f}}(X,v,t) + a \varphi_0 \mc{h}(v) \left[\bar{v}\partial_X \tilde{\varrho}(X,t) +\varrho_0 \partial_X \tilde{\mc{u}}(X,t) \right] \notag \\
     &= \frac{1}{2}\partial_X \left[\mc{D}(v)\partial_X\tilde{\mc{f}}(X,v,t) 
     -a^2 \varphi_0 \int dw |v-w| \mc{h}(w) \partial_X \tilde{\mc{f}}(X,w,t)\right], \label{BS-HD}\\
     \text{where}~&~   \tilde{\varrho}(X,t) = \int dv~\tilde{\mc{f}}(X,v,t),
     ~~\varrho_0 \tilde{\mc{u}}(X,t) = \int dv ~v \tilde{\mc{f}}(X,v,t). 
\end{align}
We note that the form of both the Euler part in Eq.~\eqref{BS-HD} [terms on the left-hand side] as well as the dissipation  [terms on the right-hand side] are different from the forms in Eq.~\eqref{eq:fhd_eq_fxvt2}. In particular, the third and fourth terms on the left-hand side of Eq.~\eqref{BS-HD}) and the second term on the right-hand side are missing from   Eq.~\eqref{eq:fhd_eq_fxvt2}. In this context, we make the following remarks:
\begin{itemize}
\item For a generic initial state with average PSD $\bar{\mc{f}}(X,v,0)$, the average current $\bar{\mc{j}}(X,v,t) = \langle j_{\rm micro}(X,v,t) \rangle$ at a phase space point $(X,v)$ at time $t$ generally has the following gradient expansion 
\cite{doyonlecturenotes}
\begin{align}
\bar{\mc{j}}(X,v,t) &\approx J_{\rm Euler}[\bar{\mc{f}}(t),X,v] + \int du ~\msc{D}[\bar{\mc{f}}(t),X,v,u]\partial_X \tilde{\mc{f}}(X,u,t), \label{j-expn} \\
\text{where}~~~~
J_{\rm Euler}[\bar{\mc{f}}(t)] &= v_{\rm eff}[\bar{\mc{f}},X,v]\bar{\mc{f}}(X,v,t), \label{j_eu} \\
\text{with}~~~~v_{\rm eff}[\bar{f}(t),X,v] &=\frac{v-\int du ~u \bar{\mc{f}}(X,u,t)}{1-\bar{\varrho}(X,t)}, \label{v_eff-func}
\end{align}
and $\msc{D}$ is related to the space-time correlation of density fluctuations \cite{doyon2023ballistic, hubner2024diffusive}. Approximating the statistical state at time $t$ to a local-equilibrium state and expanding
the above form of the current to linear order in the deviation
$\bar{\mc{f}}(X,v,t)=\varrho_0 \mc{h}(v) + \tilde{\mc{f}}(X,v,t)$ one obtains the linearized evolution equation for $\tilde{\mc{f}}$ in Eq.~\eqref{BS-HD} \cite{ percus1969exact, bulchandani2024revised}. In our phenomenological derivation starting from the model dynamics in Eq.~\eqref{Brownian-QP}, we essentially neglected the gradient contribution to the local current as present in Eq.~\eqref{j-expn}. Furthermore, we have approximated the effective velocity $v_{\rm eff}$ in Eq.~\eqref{v_eff-func} simply by $v_{\rm eff} = \frac{v_0}{1-a\varrho_0}$ --- thus neglected essential contribution even at linear order. This leads us to conclude that the model dynamics in Eq.~\eqref{Brownian-QP} is an  approximation of the microscopic dynamics of a quasiparticle given in Eq.~\eqref{eq:1quasiparticle_Xt}, that does not include contributions from gradients, necessary for the fluctuating hydrodynamic description.

\item  Ferrari and Olla~\cite{ferrari2023macroscopic} work in the Euler frame and their fluctuating hydrodynamic equations do not include the second diffusive contribution in Eq.~\eqref{BS-HD}. It can be shown (see Appendix \ref{diag-approx}) that neglecting the second term in the right-hand side of Eq.~ \eqref{BS-HD} corresponds to a relaxation-time approximation in the computation of equilibrium correlation functions using kinetic theory.  This approximation leads to a small correction to correlation functions and, consequently, it is expected that there is a negligible violation of number conservation.
\end{itemize}

\noindent
Our study can be extended in several directions. Our results for the quasiparticle statistics correspond to a special type of initial conditions, which are chosen first from factorised distributions in point particle coordinates and then converted to hard rod coordinates while keeping the number of rods on the left of the tagged quasiparticles fixed. It would be interesting to see how the statistical properties of a quasiparticle change if this condition is relaxed.
Another natural direction to explore is to study the quasiparticle motion in other integrable systems, such as Toda chain. Although some progress in this direction has already been made recently, by clearly identifying quasiparticles~\cite{aggarwal2025a,aggarwal2025b}, their fluctuations are still not fully characterized. 

\section*{Acknowledgement }
The authors thank Mrinal J Powdel for insightful discussions. AK acknowledges the financial support under projects  CRG/2021/002455 and MTR/2021/000350 from the ANRF (SERB), DST, Government of India. AD acknowledges the J.C. Bose Fellowship (JCB/2022/000014) of the Science and Engineering Research Board of the Department of Science and Technology, Government of India. HS thanks the VAJRA faculty scheme (No. VJR/2019/000079) from the Science and Engineering Research Board
(SERB), Department of Science and Technology, Government of India. SC, IM, AD, and AK would also like to acknowledge the support from the DAE, Government of India, under Project No. RTI4001. AD and AK also acknowledge the research support from the International Research Project (IRP) titled ``Classical and quantum dynamics in out of equilibrium systems" by CNRS, France. AD, AK and HS acknowledge discussions held during the program `Hydrodynamics of low-dimensional interacting systems' at the Yukawa Institute for Theoretical Physics, Kyoto,   during June 2-13, 2025.

\begin{appendix}
\numberwithin{equation}{section}
\numberwithin{figure}{section}
\section{Dynamics of single quasiparticle and two-time correlation}
\label{sec:appA}
In this appendix, we revisit the dynamics of a single quasiparticle evolving within a background of other hard rods and provide details of the derivation of the results presented in Sec.~\ref{sec:3}. Recall that the quasiparticle is defined as a special rod initially tagged with a fixed velocity $v_0$ at $t=0$ at location $X(0)=0$. Consequently, at any later time, the quasiparticle is identified as the rod having the same velocity $v_0$, which may differ from the initial tagged rod. As the system evolves, the quasiparticle moves ballistically between successive collisions with other rods. Each collision results in a positional shift of the quasiparticle, equivalent to its length $a$, and its label is transferred to the colliding rod as they exchange velocities, which can be understood from Fig.~\ref{Fig:quasiparticle_def}. Lebowitz, Percus, and Sykes \cite{lebowitz1968time} first investigated this problem for a gas of hard rods in equilibrium, demonstrating that the distribution of the position of the quasiparticle approaches a Gaussian at late times, with a diffusion constant dependent on the velocity distribution of the background particles. The same result has been reproduced using a hydrodynamic approach \cite{singh2024thermalization} by solving the (linearized) Navier-Stokes equation with identical initial conditions as for LPS. More recently, a microscopic derivation has been provided, extending the analysis to annealed and quenched initial conditions, characterized by inhomogeneous mass density profiles \cite{Mrinal_2024_HR}. We begin with a brief review of the distribution of the position of the quasiparticle at time $t$, as derived in \cite{Mrinal_2024_HR}  using the microscopic approach. Subsequently, we focus on the primary objective of this section, which is to calculate the two-time correlation function for the position of this quasiparticle.

As specified before Eq.~\eqref{eq:1quasiparticle_Xt} in sec.~\ref{sec:3}, we assume that initially the tagged rod with velocity $v_0$ is placed at the origin $X(0)=0$, while $N$ background rods on the left and right sides of it are randomly distributed over ranges $[-L,0]$ and $[0,L]$, respectively. The velocities of the background rods are chosen independently and identically from the distribution $\mc{h}(v)$. From this initial configuration, the displacement of the quasiparticle at time $t$, starting from the origin, is given by 
\begin{equation}\label{eq:1quasiparticle_XtApp}
X(t)=v_0t+a[n_{r\ell}(t)-n_{\ell r}(t)],    
\end{equation}
where $n_{r\ell}(t)$ and $n_{\ell r} (t)$ are the number of rods that collided with the quasiparticle from the right and left, respectively, up to time $t$. For fluctuating initial configurations of the background rods, these numbers also fluctuate.  To determine the distribution of $X(t)$ at time $t$, one is required to obtain the joint distribution of  $n_{r\ell}(t)$ and $n_{\ell r}(t)$. This can be calculated by mapping the motion of hard rods onto an equivalent system of hard-point particles following Eq. \eqref{eq:map-hr-hp}. In this representation, there is a corresponding (velocity) tagged point particle which also undergoes exactly $n_{r\ell}(t)$ and $n_{\ell r}(t)$ collisions from the right and left, respectively. Denoting  the joint probability distribution by ${\cal P}(n_{r\ell}, n_{\ell r},t)$,  one can formally write the distribution of $X$ at time $t$ for the quasiparticle (tagged rod) as \cite{Mrinal_2024_HR}
\begin{equation}\label{eq:P_XT_1quasiparticle}
    \mathbb{P}(X,t) = \sum_{n_{r\ell}}  \sum_{n_{\ell r}}  \mathcal{P}(n_{r\ell}, n_{\ell r},t)~
        \delta \big( X-v_0t-a\{n_{r\ell}-n_{\ell r}\} \big).
\end{equation}
 At time $t$, the tagged point particle moves to position $v_0t$, while the other particles, having random velocities, will reach different random positions. For a point particle with
velocity chosen from the distribution $\mc{h}(v)$, the single particle propagator to reach $y$ at time $t$ starting from $\bar{x}$ is given by
\begin{equation}\label{single particle propagator}
    \mc{g}(y,t|\bar{x},0) =\int_{-\infty}^{\infty}dv~\delta(y-\bar{x}-vt)~\mc{h}(v)=\frac{1}{t}\mc{h}\Bigg(\frac{y-\bar{x}}{t}\Bigg).
\end{equation}
Thus, the probability that a particle, starting from $\bar{x}$, can be found below $z$ at time $t$ is 
\begin{equation}\label{eq:barg_<}
    \mc{g}_<(z,t|\bar{x},0)=\int_{-\infty}^z dy~\mc{g}(y,t|\bar{x},0),
\end{equation}
and the probability of finding it above $z$ at time $t$ is
 \begin{equation} \label{eq:barg_>}
\mc{g}_>(z,t|\bar{x},0) = \int_z^{\infty} dy~\mc{g}(y,t|\bar{x},0).
\end{equation}
Note $ \mc{g}_<(z,t|\bar{x},0)+ \mc{g}_>(z,t|\bar{x},0)=1$ as it should be.

To compute the joint probability  $\mathcal{P}(n,m,t)$, one can imagine the motion of hard-point particles as of non-interacting particles similar to Jepsen mapping \cite{Jepsen_tagged1965}. In this case, since the particles on the left of the quasiparticle do not interact with those on the right, the joint probability $\mathcal{P}(n,m,t)$ can be written as $\mathcal{P}(n,m,t) = \mc{P}_r(n,t) \mc{P}_\ell(m,t)$ where $\mc{P}_r(n,t)$ is simply the binomial distribution of choosing $n$ out of the $N$ particles initially on the right to reach on the left of the quasiparticle at time $t$. Hence we have
\begin{flalign}
\mathcal{P}(n,m,t) &= \binom{N}{n}~ \left[\frac{p_{r\ell}}{N}\right]^n\left[\frac{p_{rr}}{N}\right]^{(N-n)}\times~\binom{N}{m}~ \left[\frac{p_{\ell r}}{N}\right]^m\left[\frac{p_{\ell \ell}}{N}\right]^{(N-m)}, 
\label{eq:mcalP(n,m,t)-an} 
\end{flalign}
where
\begin{align}
\begin{split}
&p_{r\ell}(t) = \int_0^\infty d\bar{x}~\mc{g}_{<}(v_0t,t|\bar{x},0)~\varphi_r(\bar{x}),~~ 
p_{rr}(t) 
=N - p_{r\ell}, \\
&p_{\ell \ell}(t) = \int_{-\infty}^0 d\bar{x}~\mc{g}_{<}(v_0t,t|\bar{x},0)~\varphi_\ell(\bar{x}),~~
p_{\ell r}(t) 
=N - p_{\ell \ell}.
\end{split}
\label{eq:1quasiparticle_part-probs} 
\end{align} 
Note $\frac{p_{r \ell}(t)}{N}$ represents the probability that a particle starting on the right of the quasiparticle ends up on the left of it at time $t$. A similar interpretation holds for the other probabilities.

Given $\mathcal{P}(n,m,t)$, the moment generating function (MGF) of the displacement of the quasiparticle displacement $X(t)$ can be determined as $Z(q,t)=\langle e^{-\mc{i} qX(t)}\rangle.$ Using this function and following Eq. \eqref{eq:P_XT_1quasiparticle}, the exact expression for $\mathbb{P}(X,t)$ can be derived \cite{Mrinal_2024_HR}. Furthermore, in the limit of large $N$, the authors in Ref. \cite{Mrinal_2024_HR} demonstrated that the typical distribution of the position of the quasiparticle follows a Gaussian form
\begin{equation}\label{eq:P(Z)-gauss}
    \mathbb{P}(X,t) = \frac{1}{\sqrt{2 \pi \langle X^2\rangle_c}}~\exp\left( - \frac{(X - \langle X \rangle)^2}{2 \langle X^2\rangle_c}\right),
\end{equation}
characterized by the mean and variance as given in Eq.~\eqref{mean-var-Z}. 
For the homogeneous density of the background rods, say $\varrho_0$, which gets transformed to uniform density $\varphi_0 =\frac{\varrho_0}{1-a\varrho_0}$ for point particles, the mean and the variance takes the explicit forms given in Eq.~\eqref{mean-var-Z-homo}.

The Gaussian distribution in Eq.~\eqref{eq:P(Z)-gauss} with variance growing linearly with $t$ in Eq.~\eqref{mean-var-Z-homo} seems to suggest that the quasiparticle effectively moves like a Brownian particle. In order to get more evidence on this anticipation, next, we look at the correlation of its positions at two different times.

\subsection{Two-time correlation function of single quasiparticle}
\label{auto-corr-st} 
As in the previous section, the positions of the quasiparticle at time $t_1$ and $t_2$ can be written in terms of the number of collisions it has experienced till time $t_1$ and $t_2$, respectively. Similar to Eq.~\eqref{eq:1quasiparticle_XtApp} one writes  
\begin{align}
\begin{split}
X(t_1)&=v_0t_1+a[n_{r\ell}(t_1)-n_{\ell r}(t_1)],   \\
X(t_2)&=v_0t_2+a[n_{r\ell}(t_2)-n_{\ell r}(t_2)],
\end{split}
 \label{eq:1quasiparticle_Xt12}
\end{align}
where we assumed the quasiparticle starts at the origin with velocity $v_0$. 
 Initial configurations of the background rods, $N$ on the left and $N$ on the right, are chosen in the same way as in the previous section. 
Without any loss of generality, we assume $t_2>t_1$.

Since the collisions giving rise to the jumps in the displacement appearing till time $t_1$ also contribute to the displacement at time $t_2>t_1$, the positions at the two times get correlated. More elaborately, the quasiparticle undergoes additional $\Delta n_{r\ell}(t_2)=[n_{r\ell}(t_2)-n_{r\ell}(t_1)]$ collisions from the right and  $\Delta n_{\ell r}(t_2)=[n_{\ell r}(t_2)-n_{\ell r}(t_1)]$ collisions from its left to reach $X(t_2)$ at time $t_2$.  

We proceed by defining the MGF ${\cal Z}(q_1,q_2,t_1,t_2)=\langle e^{-\mc{i} q_1 X(t_1)} e^{-\mc{i} q_2 X(t_2)}\rangle $. Inserting the expressions of $X(t_1)$ and $X(t_2)$ from Eq.~\eqref{eq:1quasiparticle_Xt12}, $\mc{Z}$ can be written as 
\begin{equation}\label{eq:tcorr_1quasiparticle_MGF}
{\cal Z}(q_1,q_2,t_1,t_2)=e^{-\mc{i} q_1v_0t_1}~e^{-\mc{i} q_2v_0t_2}~ \mathfrak{Z}(q_1,q_2,t_1,t_2),
\end{equation}
with 
\begin{align}\label{eq:tcorr_1quasiparticle_MGF_add}
\mathfrak{Z}(q_1,q_2,t_1,t_2)=\sum_{n=0}^N\sum_{\Delta n=0}^{N-n} \sum_{m=0}^N\sum_{\Delta m=0}^{N-m}&e^{-\mc{i}(q_1+q_2)an}  ~e^{-\mc{i} q_2 a\Delta n} e^{\mc{i}(q_1+q_2)am}  ~e^{\mc{i} q_2 a\Delta m} \nonumber\\& \times~\mathcal{P}(\Delta n,\Delta m,t_2; n,m,t_1),
\end{align}
where $\mathcal{P}(\Delta n,\Delta m,t_2; n,m,t_1)$ represents the joint distribution of $\Delta n_{r\ell}(t_2),~\Delta n_{\ell r}(t_2),~n_{r \ell}(t_1),$ and $n_{\ell r}(t_1)$. Once again, it is convenient to compute this probability considering the dynamics of the corresponding point particles. 
It is easy to see that, due to the ballistic motion of the particles, the additional number of collisions, $\Delta n$, experienced by the quasiparticle from the right during the interval $[t_1, t_2]$ must originate solely from the remaining $(N - n)$ particles that were still to the right of the quasiparticle at time $t_1$.
Similarly, $\Delta m$ collisions must originate from $(N-m)$ particles still on the left of the quasiparticle at time $t_1$. Hence, one can write 
\begin{align}
  \mathcal{P}(\Delta n,\Delta m,t_2; n,m,t_1)=   \mathcal{P}_\ell(\Delta n,t_2; n,t_1) \times \mathcal{P}_r(\Delta m,t_2; m,t_1),
\end{align}
where $\mc{P}_\ell(\Delta n,t_2; n,t_1)$ represents the probability that the quasiparticle faces $n$ and $\Delta n$ collisions from right in time durations $t_1$ and $t_2-t_1$, respectively. Similar interpretation holds for $\mc{P}_r(\Delta m,t_2; m,t_1)$. It is easy to show that 
\begingroup\makeatletter\def\f@size{11}\check@mathfonts
\def\maketag@@@#1{\hbox{\m@th\large\normalfont#1}}%
\begin{align}
\mc{P}_\ell(\Delta n,t_2; n,t_1)=\left[\binom{N}{n}~ \left[\frac{p_{r\ell}}{N}\right]^n\left[\frac{p_{rr}}{N}\right]^{(N-n)}\right] \times \left[\binom{N-n}{\Delta n}~ \left[q_{rr\ell}\right]^{\Delta n}\left[q_{rrr}\right]^{(N-n-\Delta n)}\right], \label{mcP-ell}
\end{align}
\endgroup
where $p_{r \ell}$ and $p_{rr}$ are given in Eq.~\eqref{eq:1quasiparticle_part-probs} and, 
\begin{align}
q_{rr\ell}= \frac{1}{ p_{rr}}~\int_{0}^{\infty}d\bar x\int_{v_0t_1}^{\infty}dx_1\int_{-\infty}^{v_0t_2}dx_2~\mc{g}(x_2,t_2|x_1,t_1)~\mc{g}(x_1,t_1|\bar{x},0)~\varphi_r(\bar x),
\end{align}
and $q_{rrr}=1-q_{rr\ell}$. Here, $q_{rr\ell}$ represents the probability that a particle starting on the right of the quasiparticle reaches a location on the left of it at time $t_2$ given that it was on the right at an earlier time $t_1$. Hence, the first factor in Eq.~\eqref{mcP-ell} represents the probability that $n$ out of $N$ particles initially on the right of the quasiparticle move to the left of it at time $t_1$. The second factor represents the conditional probability that out of the remaining $N-n$ particles on the right at time $t_1$, $\Delta n$ particles move to the left of the quasiparticle in duration $t_2-t_1$. Similarly, one can write 
\begingroup\makeatletter\def\f@size{11}\check@mathfonts
\def\maketag@@@#1{\hbox{\m@th\large\normalfont#1}}%
\begin{align}
\mc{P}_r(\Delta m,t_2; m,t_1)=\binom{N}{m}~ \left[\frac{p_{\ell r}}{N}\right]^m\left[\frac{p_{\ell \ell}}{N}\right]^{(N-m)}\times\binom{N-m}{\Delta m}~ \left[q_{\ell \ell r}\right]^{\Delta m}\left[q_{\ell \ell \ell}\right]^{(N-m-\Delta m)},\label{mcP-r}
\end{align}
\endgroup
with
\begin{align}
q_{\ell \ell r}= \frac{1}{ p_{\ell \ell}}~\int_{-\infty}^{0}d\Bar{x} \int_{-\infty}^{v_0t_1}dx_1\int_{v_0t_2}^{\infty}dx_2~\mc{g}(x_2,t_2|x_1,t_1)~\mc{g}(x_1,t_1|\bar{x},0) ~\varphi_{\ell}(\bar x),
\end{align}
and $q_{\ell \ell \ell}=1-q_{\ell \ell r}$. 
Using the forms of $\mc{P}_\ell$ and $\mc{P}_r$ from Eqs.~\eqref{mcP-ell} and \eqref{mcP-r} in Eq.~\eqref{eq:tcorr_1quasiparticle_MGF_add} and performing the sums, we get 
\begin{equation}\label{eq:MGF_1T_different_t}
{\cal Z}(q_2,t_2,q_1,t_1)=e^{-\mc{i} q_1v_0t_1}~e^{-\mc{i} q_2v_0t_2} \mathfrak{Z}_{r\ell}(q_1,q_2,t_1,t_2)~ \mathfrak{Z}_{\ell r}(-q_1,-q_2,t_1,t_2),
\end{equation}
with,
\begin{align}
\begin{split}
  \mathfrak{Z}_{r\ell}(q_1,q_2,t)
  =\Bigg[1+\frac{p_{r\ell}}{N}\Big(e^{-\mc{i}(q_1+q_2)a}-e^{-\mc{i} q_2 a}\Big)+\frac{\bar{p}_{r\ell}}{N}\Big(e^{-\mc{i} q_2 a}-1\Big)\Bigg]^N, \\
  \mathfrak{Z}_{\ell r}(-q_1,-q_2,t)
    =\Bigg[1+\frac{p_{\ell r}}{N}\Big(e^{\mc{i}(q_1+q_2)a}-e^{\mc{i} q_2 a}\Big)+\frac{\bar{p}_{\ell r}}{N}\Big(e^{\mc{i} q_2 a}-1\Big)\Bigg]^N,
    \end{split}
\end{align}
where $p_{r\ell}, p_{\ell r}$ are given in Eq. \eqref{eq:1quasiparticle_part-probs}, and 
\begin{align}
   &\frac{\bar{p}_{r\ell}}{N}=\frac{p_{rr}~q_{rr\ell}}{N}+\frac{p_{r\ell}}{N}=1- \frac{\bar{p}_{rr}}{N},~~\frac{\bar{p}_{\ell r}}{N}=\frac{p_{\ell \ell}~q_{\ell \ell r}}{N}+\frac{p_{\ell r}}{N}=1-\frac{\bar{p}_{\ell \ell}}{N}.
\label{eq:2time_corr_1tarcer_probs}
\end{align}
In the limit $N \to \infty$, the expression of the MGF as in Eq. \eqref{eq:MGF_1T_different_t} becomes
\begin{align}
    {\cal Z}(q_2,t_2,q_1,t_1)&=e^{-\mc{i} q_1v_0t_1}~e^{-\mc{i} q_2v_0t_2}~\exp\Big[p_{r\ell}(e^{-\mc{i}(q_1+q_2)a}-e^{-\mc{i} q_2 a})+\bar{p}_{r\ell }(e^{-\mc{i} q_2 a}-1)\Big ]\nonumber\\&
    ~~~~\times\exp\Big[p_{\ell r}(e^{\mc{i}(q_1+q_2)a}-e^{\mc{i} q_2 a})+\bar{p}_{\ell r}(e^{\mc{i} q_2 a}-1) \Big].
\end{align}
Expanding the $e^{\pm \mc{i} q_{1,2}a}$ terms up to quadratic order in $q_{1,2}$ we find 
\begin{align}
    {\cal Z}(q_2,t_2,q_1,t_1)&=\exp\Bigg( -\mc{i} q_1\big[v_0t_1+a(p_{rl}-p_{lr})\big]-\mc{i} q_2\big[v_0t_2+a(\bar p_{rl}-\bar p_{lr})\big]\Bigg)\nonumber \\&
    ~~\times \exp \Bigg( -\frac{q_1^2a^2}{2}\big[p_{rl}+p_{lr}\big]-\frac{q_2^2a^2}{2}\big[\bar p_{rl}+\bar p_{lr}\big] -q_1q_2a^2\big[p_{rl}+p_{lr}\big]  \Bigg).
\end{align}
Consequently, the average values of the displacements $X(t_1)$ and $X(t_2)$ comes out to be
\begin{align}\label{eq:2timecorr_mean}
    &\langle X(t_1)\rangle=v_0t_1+a(p_{rl}-p_{lr}),~\langle X(t_2)\rangle=v_0t_2+a(\bar p_{rl}-\bar p_{lr}),
\end{align}
whereas their variances appear to be 
\begin{align}\label{eq:2timecorr_var}
    &\langle X^2(t_1)\rangle=a(p_{rl}+ p_{lr}), ~\langle X^2(t_2)\rangle=a^2(\bar p_{rl}+\bar p_{lr}).
\end{align}
Taking derivatives with respect to both $q_1,q_2$, one gets  the two-time correlation function as 
\begin{align}\label{eq:2timecorr_covar}
 \langle X(t_1)X(t_2)\rangle_c=-\frac{d^2\ln {\cal Z}(q_2,t_2,q_1,t_1)}{dq_1dq_2} \bigg|_{q_1=0,q_2=0}=a^2(p_{rl}+p_{lr}),
\end{align}
as in Eq.~\eqref{1t:acorr}.

\section{Derivation of Eqs.~\eqref{eq:2quasiparticle_separate_mean_gn} - \eqref{eq:2quasiparticle_separate_cov_gn}}
\label{app:2-Trc-calc}

\begin{figure}[t]
\centering
\includegraphics[width= 3.8 in]{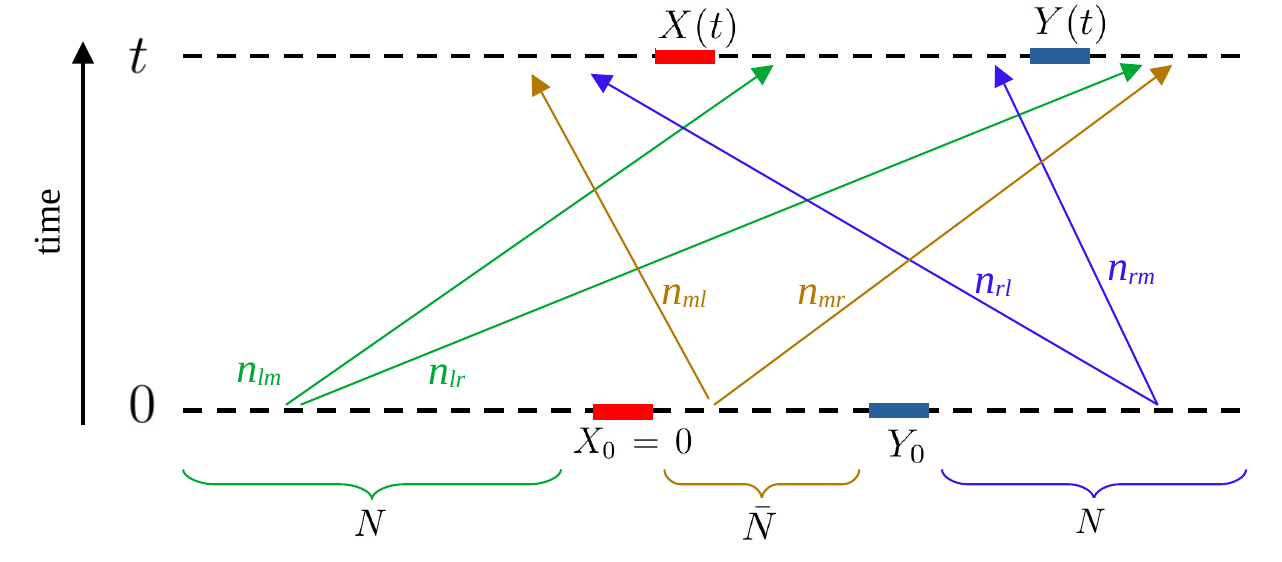} 
\caption{Schematic diagram illustrating the evolution of two quasiparticles, marked as red and blue, moving in a system of $\msc{N}=(2N+\bar N)$ hard rods. These quasiparticles are initially positioned at $X(0)=X_0 =0$ and $Y(0)=Y_0>0$ with velocities $v_0$ and $u_0>v_0$, respectively. The system at $t=0,$ consists of $N$ rods to the left of $X_0$, $N$ rods to the right of $Y_0$, and $\bar N$ rods between $X_0$ and $Y_0$. Here $n_{r\ell}~(n_{\ell r})$ represents the number of rods that were initially to the right (left) of $Y_0$ ($X_0$) and then collided with both $X(t)$ and $Y(t)$ during the time interval $t$. $n_{r m}$ ($n_{\ell m}$) be the number of rods that were initially to the right (left) of $Y_0$ ($X_0$), that collided only with $Y(t)$ ($X(t)$) during time $t$, and similarly $n_{m \ell} ~(n_{mr})$ depicts the number of rods that were initially present between $X_0$ and $Y_0$ and which collided with $X(t), ~(Y(t))$ during time $t$.}
\label{Fig:2quasiparticle_separated}
\end{figure}

\noindent
We start by  defining the MGF 
\begin{align}
&~~~~~~{\cal Z}(q_1,q_2,t)=\langle e^{-\mc{i} q_1 X(t)} e^{-\mc{i} q_2 Y(t)}\rangle =e^{-\mc{i} q_1v_0t}~e^{-\mc{i} q_2(Y_0+u_0t)}~ \mathfrak{Z}(q_1,q_2,t), \cr
&\text{where}, \label{eq:2quasiparticle_origin_MGF} \\ 
&~~~~~~\mathfrak{Z}(q_1,q_2,t)=\left \langle e^{-\mc{i} q_1 a [n_{r\ell}(t)+n_{m\ell}(t)-n_{\ell m}(t)-n_{\ell r}(t)]}~e^{-\mc{i} q_2a [n_{rm}(t)+n_{r\ell}(t)-n_{mr}(t)-n_{\ell r}(t)]} \right \rangle. \notag
\end{align}
This $\mathfrak{Z}(q_1,q_2,t)$ can be written explicitly as  
 \begin{align}\label{eq:tcorr_2quasiparticle_MGF_add}
 \begin{split}
\mathfrak{Z}(q_1,q_2,t)&= \sum_{n_{\ell r}=0}^{N} \sum_{n_{\ell m}=0}^{N-n_{\ell r}} \sum_{n_{r\ell}=0}^{N}\sum_{n_{rm}=0}^{N-n_{r\ell}}  \sum_{n_{m\ell}=0}^{\bar{N}}\sum_{n_{mr}=0}^{\bar{N}-n_{m\ell}}  ~
{\cal{P}}(n_{rm},n_{r\ell},n_{m\ell},n_{mr},n_{\ell m},n_{\ell r},t) \cr
& \times~e^{-\mc{i} q_1 a [n_{r\ell}(t)+n_{m\ell}(t)-n_{\ell m}(t)-n_{\ell r}(t)]}~e^{-\mc{i} q_2 a [n_{rm}(t)+n_{r\ell}(t)-n_{mr}(t)-n_{\ell r}(t)]},  
\end{split}
\end{align}
where $ {\cal{P}}(n_{rm},n_{r\ell},n_{m\ell},n_{mr},n_{\ell m},n_{\ell r},t)$ denotes the joint distribution of the number of different collisions (see Fig.~\ref{Fig:2quasiparticle_separated}).
To compute $\mathfrak{Z}(q_1,q_2,t)$, one requires an expression of this joint probability. Following similar arguments presented in Sec.~\ref{sec:3}, one can realise that this distribution has the following structure
\begin{equation}\label{eq:2quasiparticle_separated_prob}
    {\cal{P}}(n_{rm},n_{r\ell},n_{m\ell},n_{mr},n_{\ell m},n_{\ell r},t)={\cal{P}}_r(n_{rm},n_{r\ell},t) {\cal{P}}_m(n_{m\ell},n_{mr},t) {\cal{P}}_l(n_{\ell m},n_{\ell r},t),
\end{equation}
because the initial statistics of the particles on the left of the position of the quasiparticle at $X_0$, right of the position of the quasiparticle $Y_0$, and in between them are essentially independent of each other.
The probabilities $\mc{P}_r,~\mc{P}_\ell$ and $\mc{P}_m$ are given explicitly as 
\begingroup\makeatletter\def\f@size{11}\check@mathfonts
\def\maketag@@@#1{\hbox{\m@th\large\normalfont#1}}%
\begin{align}
        & {\cal{P}}_r(n_{rm},n_{r\ell},t) ={N \choose n_{r\ell}} {N-n_{r\ell} \choose n_{rm} } \left[\frac{p_{r\ell}}{N}\right]^{n_{r\ell}}\left[\frac{p_{rm}}{N}\right]^{n_{rm}}\left[1-\frac{p_{rm}+p_{r\ell}}{N}\right]^{N-n_{rm}-n_{r\ell}}, \cr
   &{\cal{P}}_m(n_{m\ell},n_{mr},t)={\Bar{N} \choose n_{m\ell}} {\bar{N}-n_{m\ell} \choose n_{mr} } \left[\frac{p_{m\ell}}{\Bar{N}}\right]^{n_{m\ell}}\left[\frac{p_{mr}}{\Bar{N}}\right]^{n_{mr}}\left[1-\frac{p_{m\ell}+p_{mr}}{\bar{N}}\right]^{\bar{N}-n_{m\ell}-n_{mr}}, \cr 
    &{\cal{P}}_l(n_{\ell m},n_{\ell r},t)={N \choose n_{\ell r}} {N-n_{\ell r} \choose n_{\ell m} } \left[\frac{p_{\ell r}}{N}\right]^{n_{\ell r}}\left[\frac{p_{\ell m}}{N}\right]^{n_{\ell m}}\left[1-\frac{p_{\ell m}+p_{\ell r}}{N}\right]^{N-n_{\ell m}-n_{\ell r}}, \label{eq:2quasiparticle_separated_prob1}
\end{align}
\endgroup
where
\begin{align}\label{eq:plR plM prL prM 2T at 0} 
    \begin{split}
        p_{\ell r}(t) &= \int_{-\infty}^0 d \bar{x} ~ \mc{g}_>(y_0+u_0t,t|\bar{x},0)~\varphi_l(\bar{x}), \cr
        p_{\ell m}(t) &= \int_{-\infty}^0 d \bar{x} ~ \mc{g}_m(v_0t,y_0+u_0t,t|\bar{x},0)~\varphi_l(\bar{x}), \cr 
        p_{r\ell}(t) &= \int_{y_0}^{\infty} d \bar{x} ~ \mc{g}_<(v_0t,t|\bar{x},0)~\varphi_r(\bar{x}),\cr
        p_{rm}(t) &= \int_{y_0}^{\infty} d \bar{x} ~ \mc{g}_m(v_0t,y_0+u_0t,t|\bar{x},0)~\varphi_r(\bar{x}),
    \end{split}
\end{align}
with $\mc{g}_m(z,x,t|\bar{x},0) = \int_{z}^x dy ~ ~  \mc{g}(y,t|{\bar{x}},0) = \mc{g}_<(x,t|\bar{x},0) - \mc{g}_<(z,t|\bar{x},0)$ and, 
\begin{align}\label{eq:pmR pmL}
    \begin{split}
        p_{mr}(t)&=\int_{0}^{y_0}d{\bar{x}}~\mc{g}_>(y_0+u_0t,t|\bar{x},0)\varphi_{m}({\bar{x}}), \cr 
        p_{m\ell}(t)&=\int_{0}^{y_0}d{\bar{x}}~\mc{g}_<(v_0t,t|\bar{x},0)\varphi_{m}({\bar{x}}),
    \end{split}
\end{align}
with $y_0=Y_0-(\bar{N}+1)a$, which is the separation of quasiparticles in the point particle picture. Note the function \( \mc{g}_m(z, x, t | \bar{x}, 0) \) represents the probability that a particle, initially at \( \bar{x} \) at time \( t = 0 \), is found in the range $[z,x]$. Performing the sums in Eq.~\eqref{eq:tcorr_2quasiparticle_MGF_add} one finds explicit expression of $\mathfrak{Z}(q_1,q_2,t)$, using which in Eq.~\eqref{eq:2quasiparticle_origin_MGF}, one gets 
\begin{gather}\label{eq:2quasiparticle_separated_MGF}
    \begin{split}
        {\cal Z}(q_1,q_2,t) & =e^{-\mc{i} q_1v_0t}~ e^{-\mc{i} q_2(Y_0+u_0t)}~ \left[1+\left(e^{-\mc{i} q_1a}-1\right)\frac{p_{m\ell}}{\bar N}+\left(e^{\mc{i} q_2a}-1\right)\frac{p_{mr}}{\bar N}\right]^{\bar N}
   \\& \hspace{2cm} \times \left[1+\left(e^{-\mc{i} a(q_1+q_2)}-1\right)\frac{p_{r\ell}}{N} +\left(e^{-\mc{i} aq_2}-1\right)\frac{p_{rm}}{N}\right]^N \\& \hspace{2cm} \times \left[1+\left(e^{\mc{i} a(q_1+q_2)}-1\right)\frac{p_{\ell r}}{N} +\left(e^{\mc{i} aq_1}-1\right)\frac{p_{\ell m}}{N}\right]^N.
    \end{split}
\end{gather}
For finite $\bar{N}$, we take the limit of large $N$ and $L$ keeping the ratio finite. Expanding the exponent to quadratic order in $(qa)$, we get the following approximate expression
\begin{gather}
    \begin{split}
        {\cal Z}(q_1, q_2, t) & \approx \exp \Bigg[
-\mc{i} q_1 \Big( v_0 t + a (p_{r\ell} - p_{\ell r} - p_{\ell m}) \Big)
-\mc{i} q_2 \Big(Y_0+u_0 t + a (p_{r\ell} + p_{rm} - p_{\ell r}) \Big) 
\\&  -\frac{q_1^2}{2} a^2 \big( p_{r\ell} + p_{\ell m} + p_{\ell r} \big)
-\frac{q_2^2}{2} a^2 \big( p_{r\ell } + p_{rm} + p_{\ell r} \big)
-q_1 q_2 a^2 \big( p_{r\ell } + p_{\ell r} \big)
\Bigg]   \\& ~~~~~~ \times\Bigg[1-\Big( \mc{i} q_1a+\frac{q_1^2a^2}{2}\Big)\frac{p_{m\ell}}{\bar N}+\Big(\mc{i} q_2a-\frac{q_2^2a^2}{2}\Big)\frac{p_{mr}}{\bar N}\Bigg]^{\bar N}, \label{MGF for 2T separated}
    \end{split}
\end{gather}
where the functions $p_{r\ell},~p_{\ell r}$ etc. can be defined in terms of the propagator $\mc{g}_{\lessgtr}$ as given in Eqs.~\eqref{eq:plR plM prL prM 2T at 0} and \eqref{eq:pmR pmL}. Taking derivatives of $-\ln {\cal Z}(q_1, q_2, t) $ with respect to $q_1$ and $q_2$, one finds the cumulants of $X(t)$ and $Y(t)$ given in Eqs.~\eqref{eq:2quasiparticle_separate_mean_gn} - \eqref{eq:2quasiparticle_separate_cov_gn}.

\subsection{Mean, Variance and covariance in case of domain-wall initial condition}
\label{mvc-dw}
In Sec \ref{sec:two_quasiparticles}, we obtained the exact results for the mean, variance, and covariance of the positions of two quasiparticles, which were initially positioned at $X(0)=X_0=0$ and $Y(0)=Y_0$. These results, presented in  Eqs.~\eqref{eq:2quasiparticle_separate_mean_gn} - \eqref{eq:2quasiparticle_separate_cov_gn}, hold for any arbitrary initial mass density profiles. Here, we specialize to the case of a domain-wall initial condition, where $\varphi_l$ denotes the mass density profile to the left of $X_0$, $\varphi_r$ to the right of $Y_0$, and $\varphi_m$ in the region between $X_0$ and $Y_0$. Explicit expressions for mean, variance, and covariance for this case are given as follows: 
\begin{gather}
\label{eq:mean-dw}
    \begin{split}
       & \langle X(t) \rangle = v_0t+a t \Big(\varphi_r {\msc{F}}(v_0-y_0/t)-\varphi_{\ell} \bar{{\msc{F}}}(v_0)+\varphi_{m} \Big[{\msc{F}}(v_0) - {\msc{F}}(v_0-y_0/t)  \Big]\Big),\\&
        \langle Y(t) \rangle = Y_0+u_0t+a t \Big(\varphi_r {\msc{F}}(u_0)-\varphi_{\ell} \bar{{\msc{F}}}(u_0+y_0/t)+\varphi_{m} \Big[\bar{{\msc{F}}}(u_0+y_0/t) - \bar{{\msc{F}}}(u_0)  \Big]\Big),
    \end{split}
\end{gather}
\begin{equation}
    \begin{split}
    & \langle X^2(t)\rangle_c = a^2 t \Big[ \varphi_r {\msc{F}}(v_0-y_0/t) + \varphi_{\ell} \bar{{\msc{F}}}(v_0) + \varphi_m \big({\msc{F}}(v_0) - {\msc{F}}(v_0-y_0/t) \big) \Big] \\& \hspace{5.5cm}-\frac{a^2}{\bar N}\varphi_{m}^2 t^2 \Big[{\msc{F}}(v_0)-{\msc{F}}(v_0-y_0/t)\Big]^2,
    \\&
       \langle Y^2(t)\rangle_c =  a^2 t \Big[ \varphi_r {\msc{F}}(u_0) + \varphi_{\ell} \bar{{\msc{F}}}(u_0+y_0/t) + \varphi_m \big(\bar{{\msc{F}}}(u_0) - \bar{{\msc{F}}}(u_0+y_0/t) \big) \Big] \\& \hspace{5.5cm} -\frac{a^2}{\bar N}\varphi_{m}^2 t^2 \Big[\bar{{\msc{F}}}(u_0)-\bar{{\msc{F}}}(u_0+y_0/t)\Big]^2,
    \end{split}
\end{equation}
\begin{equation}
    \begin{split}
      \langle X(t)Y(t) \rangle_c = & a^2 t \Big[ \varphi_r {\msc{F}}(v_0-y_0/t)+ \varphi_{\ell} \bar{{\msc{F}}}(u_0+y_0/t) \Big]
    \\&~~~~+\frac{a^2}{\bar N}\varphi^2_{m} t^2\Big[{\msc{F}}(v_0)-{\msc{F}}(v_0-y_0/t)\Big]\Big[\bar{{\msc{F}}}(u_0)-\bar{{\msc{F}}}(u_0+y_0/t)\Big],
    \end{split}
\end{equation}
\begin{equation}
    \begin{split}
        \langle (Y-X)^2\rangle_c = & a^2 t \Big[ \varphi_r \Big( {\msc{F}}(u_0)- {\msc{F}} (v_0-y_0/t)\Big) + \varphi_l \Big( \bar{{\msc{F}}}(v_0) - \bar{{\msc{F}}}(u_0+y_0/t) \Big)  \\&
        + \varphi_m \Big(\ {\msc{F}}(v_0)-{\msc{F}}(v_0-y_0/t) + \bar{{\msc{F}}}(u_0) - \bar{{\msc{F}}} (u_0+y_0/t)\Big)\Big]\\&
        -\frac{a^2 \varphi_m^2 t^2}{\bar{N}} \Big[  \big({\msc{F}}(v_0)-{\msc{F}}(v_0-y_0/t) \big) + \big( \bar{{\msc{F}}}(u_0) - \bar{{\msc{F}}}(u_0+y_0/t) \big) \Big]^2,
    \end{split}
    \label{eq:sep-dw}
\end{equation}
where $\msc{F}(v)$ and $\bar{\msc{F}}(v)$ are defined in Eq. \eqref{defining F Fbar of v}.
As observed in the case of a homogeneous state, the mean positions of quasiparticles in this setup grow ballistically with an effective velocity. On the other hand, linear growth of the variance of the individual quasiparticles indicates diffusive behavior. However, using Eq. \eqref{eq:sep-dw}, at large time limit, for $u_0=v_0$ ({\it i.e.} when the quasiparticles have the same velocities) the variance of the separation of the two quasiparticles becomes  
\begin{equation}
    \begin{split}
        \frac{\langle (Y-X)^2 \rangle}{t} =  a^2\Big(\frac{y_0}{t}\Big)\Bigg(&\Big[ \varphi_r \int_{-\infty}^{v_0} dw ~ \mc{h}(w) + \varphi_l \int_{v_0}^{\infty} dw ~ \mc{h}(w) \Big]\\&
        - \frac{\mc{h}(v_0)}{2} \big[\varphi_r+\varphi_l-2\varphi_m]\Big(\frac{y_0}{t}\Big) -\varphi_m \mc h^2(v_0) \Big(\frac{y_0}{t}\Big)^2\Bigg).
    \end{split}
\end{equation}
This once again indicates that the two quasiparticles are performing a rigid-body-like motion at late times.   

\section{Quasiparticle dynamics with quenched initial positions}
\label{app:2quasiparticle_quenched_prob}
In Sec. \ref{sec:two_quasiparticles}, we discussed the dynamics of two quasiparticles evolving within a background of interacting hard rods under annealed initial conditions, using a microscopic framework. In this section, we revisit the dynamics under quenched initial conditions (as described in Eq. \eqref{eq:quenched_IC_define}) and derive statistical properties of two quasiparticles, initially positioned at $X(0)=X_0=0$ and $Y(0)=Y_0 >0$ with velocities $v_0$ and $u_0$, respectively. We further assume that there are $\bar{N}$ rods between the two quasiparticles.  To compute the mean, variance, and covariance, we once again evaluate the moment generating function 
\begin{align}
&~~~~~~{\cal Z}(q_1,q_2,t)=\langle e^{-\mc{i} q_1 X(t)} e^{-\mc{i} q_2 Y(t)}\rangle =e^{-\mc{i} q_1v_0t}~e^{-\mc{i} q_2(Y_0+u_0t)}~ \mathfrak{Z}(q_1,q_2,t),
\end{align}
where,  
\begin{align}
 \begin{split}
\mathfrak{Z}(q_1,q_2,t)&= \sum_{n_{\ell r}=0}^{N} \sum_{n_{\ell m}=0}^{N-n_{\ell r}} \sum_{n_{r\ell}=0}^{N}\sum_{n_{rm}=0}^{N-n_{r\ell}}  \sum_{n_{m\ell}=0}^{\bar{N}}\sum_{n_{mr}=0}^{\bar{N}-n_{m\ell}}  ~
{\cal{P}}(n_{rm},n_{r\ell},n_{m\ell},n_{mr},n_{\ell m},n_{\ell r},t) \cr
& \times~e^{-\mc{i} q_1 a [n_{r\ell}(t)+n_{m\ell}(t)-n_{\ell m}(t)-n_{\ell r}(t)]}~e^{-\mc{i} q_2 a [n_{rm}(t)+n_{r\ell}(t)-n_{mr}(t)-n_{\ell r}(t)]},  
\end{split}
\end{align}
and ${\cal{P}}(n_{rm},n_{r\ell},n_{m\ell},n_{mr},n_{\ell m},n_{\ell r},t)$ represents the joint distribution of the number of distinct collisions experienced by the quasiparticles at time $t$. To evaluate ${\cal Z}(q_1,q_2,t)$, we begin by computing the corresponding joint distribution within the point particle representation. Below we assume $u_0>v_0$ to present the calculation of the MGF $\mc{Z}(q_1,q_2,t)$, however, the procedure can be easily extended for the opposite case as well.

Since, the dynamics of the point particles to the left of  $X_0$, to the right of $Y_0$, and those positioned between  $X_0$ and $Y_0$ are mutually independent, the joint probability $ \mathcal{P}(n_{rm}=m,n_{r\ell}=n,n_{m\ell}=s, n_{mr}=r, n_{\ell m}=\mu,n_{\ell r}=\nu, t) $ at time $t$ can be factorized as
\begin{equation}\label{eq:QIC_2quasiparticle_separated_prob_app}
 {\cal{P}}(m,n,s,r,\mu,\nu,t)={\cal{P}}_r(m,n,t) {\cal{P}}_m(s,r,t) {\cal{P}}_{\ell}(\mu,\nu,t).
\end{equation}
We assume the initial positions of the point particles, right of $Y_0$ are denoted as $\{\bar x_j\}$ (with $j=1,2,\dots,N$), those to the left of $X_0$ are represented as $\{\bar x_k\}$ ($k=1,2,\dots,N$), and the positions of the point particles between $X_0$ and $Y_0$ are denoted as $\{\bar x_l\}$ ($l=1,2,\dots,\bar N$). Now, recall in terms of the propagator as in Eq. \eqref{single particle propagator} we can write the probability that a particle, starting from $\bar{x}$, can be found below $z$ at time $t$ is $\mc{g}_<(z,t|\bar{x},0)$
and the probability of finding it above $z$ at time $t$ is $\mc{g}_>(z,t|\bar{x},0)$.
Similarly, the probability that a particle, starting from $\bar{x}$, can be found between $z$ and $x$ at time $t$ can be represented as
$\mc{g}_m(z,x,t|\bar{x},0) = \mc{g}_<(x,t|\bar{x},0) - \mc{g}_<(z,t|\bar{x},0)$.
In terms of these probabilities, one can write the distributions ${\cal{P}}_r(m,n,t), {\cal{P}}_m(s,r,t), {\cal{P}}_{\ell}(\mu,\nu,t)$ in a similar way as discussed in Ref. \cite{Mrinal_2024_HR}
\begin{gather}\label{eq:QIC_2quasiparticle_separated_prob_general}
    \begin{split}
      &{\cal{P}}_r(m,n,t)= \Bigg[\prod_{j=1}^N\Bigg(\sum_{\epsilon_j=0,1,2}\Bigg) ~g_{\epsilon_j}(t|\Bar x_j)~\delta_{\sum_{j=1}^N (\delta_{\epsilon_j,0}),n}~\delta_{\sum_{j=1}^N (\delta_{\epsilon_j,1}),m}\Bigg],\\
    &{\cal{P}}_{\ell}(\mu,\nu,t)= \Bigg[\prod_{k=1}^N\Bigg(\sum_{\epsilon_{k}=0,1,2}\Bigg) ~g_{\epsilon_{k}}(t|\Bar x_k)~\delta_{\sum_{k=1}^N (\delta_{\epsilon_{k},1}),\mu}~\delta_{\sum_{j=1}^N (\delta_{\epsilon_{k},2}),\nu}\Bigg],\\
    & {\cal{P}}_{m}(s,r,t)=
    \Bigg[\prod_{l=1}^{\bar{N}}\Bigg(\sum_{\epsilon_l=0,1,2}\Bigg) ~g_{\epsilon_l}(t|\Bar x_l)~\delta_{\sum_{l=1}^{\Bar{N}} (\delta_{\epsilon_l,0}),s}~\delta_{\sum_{l=1}^{\Bar{N}} (\delta_{\epsilon_l,2}),r}\Bigg],
    \end{split}
\end{gather}
where the probability $g_{\epsilon}(t|\Bar x)$ can be interpreted as follows: for $\epsilon =0$ it represents the probability that a particle, starting from $\bar x$, can be found below $v_0t$ at time $t$, for $\epsilon =1$ it corresponds to the probability that a particle, starting from $\bar x$, can be found between $v_0t$ and $(y_0+u_0t)$ at time $t,$ and similarly for $\epsilon =2$ it denotes the probability that a particle, starting from $\bar x$, can be found above $y_0+u_0t$ at time $t.$ This leads to the expression 
\begin{equation}\label{g epsilon}
    g_{\epsilon}(t|\Bar x)=\mc{g}_<(v_0t,t|\Bar x,0)\delta_{\epsilon,0}+\mc{g}_m(y_0+u_0t,v_0t,t|\Bar x,0)\delta_{\epsilon,1}+\mc{g}_>(y_0+u_0t,t|\Bar x,0)\delta_{\epsilon,2}.
\end{equation}
The Kronecker deltas $\delta_{\sum_{j=1}^N (\delta_{\epsilon_j,0}),n}$ and $\delta_{\sum_{j=1}^N (\delta_{\epsilon_j,1}),m}$ in the first line of Eq. \eqref{eq:QIC_2quasiparticle_separated_prob_general} respectively ensure that in the point particle representation, out of the $N$ particles initially located to the right of the origin, $n$ particles have moved to the left of the position $v_0t$ by time $t$, while $m$ particles have reached positions between $v_0 t$ and $(y_0+u_0 t)$. Similarly, the Kronecker deltas in the second line enforce that out of the $N$ particles initially located to the left of the origin, $\mu$ particles have reached positions between $v_0 t$  and $(y_0+u_0 t)$, while $\nu$ particles have right to the position $(y_0+u_0 t)$ by time $t$, and the Kronecker deltas in the third line represents that out of the $\bar N$ particles initially located between the quasiparticles, $s$ particles have reached positions left of $v_0 t$, while $r$ particles have reached positions right of $(y_0+u_0 t).$ It is worth noting that under the annealed initial condition, the distributions ${\cal{P}}_r(m,n,t), {\cal{P}}_m(s,r,t), {\cal{P}}_{\ell}(\mu,\nu,t)$ as in Eq. \eqref{eq:QIC_2quasiparticle_separated_prob_general} reduces to the form obtained in Eq. \eqref{eq:2quasiparticle_separated_prob1}. 

To proceed further, we first use the integral representation of the Kronecker delta,  
$\delta_{n,0} = \int_{-\pi}^{\pi} d\theta~e^{\mc{i} n\theta}$ 
in Eq. \eqref{eq:QIC_2quasiparticle_separated_prob_general}, and performing some simplifications, we express the equations as
\begingroup\makeatletter\def\f@size{10}\check@mathfonts
\def\maketag@@@#1{\hbox{\m@th\large\normalfont#1}}%
\begin{gather}\label{eq:2quasiparticle_separated_prob1_app}
    \begin{split}
        & {\cal P}_r(m, n, t)=\frac{1}{4\pi^2}\int_{-\pi}^{\pi}d\theta_1\int_{-\pi}^{\pi}d\theta_2 \ e^{-\mc{i} n\theta_1}e^{-\mc{i} m\theta_2}\prod_{j=1}^N\Big[1+(e^{\mc{i} \theta_1}-1)g_<(v_0t,t|\bar x_j,0) \\& \hspace{8cm}+(e^{\mc{i} \theta_2}-1)g_m(y_0+u_0t,v_0t,t|\bar x_j,0)\Big], \\&
         {\cal P}_m(s,r, t)=\frac{1}{4\pi^2}\int_{-\pi}^{\pi}d\chi_1\int_{-\pi}^{\pi}d\chi_2 \ e^{-\mc{i} s \chi_1} e^{-\mc{i} r \chi_2}\prod_{l=1}^{\bar{N}} \Big[ 1+(e^{\mc{i} \chi_1}-1)g_<(y_0+u_0t,t|\bar x_l,0)\\& \hspace{8.5cm}+(e^{\mc{i} \chi_2}-1)g_>(v_0t,t|\bar x_l,0) \Big],\\&
    {\cal P}_{\ell}(\mu, \nu, t)=\frac{1}{4\pi^2}\int_{-\pi}^{\pi}d\xi_1\int_{-\pi}^{\pi}d\xi_2 \ e^{-\mc{i} \mu\xi_1} e^{-\mc{i} \nu\xi_2}\prod_{k=1}^{N} \Big[ 1+(e^{\mc{i} \xi_1}-1)g_m(y_0+u_0t,v_0t,t|\bar x_k,0)\\& \hspace{8.5cm}+(e^{\mc{i} \xi_2}-1)g_>(y_0+u_0t,t|\bar x_k,0) \Big].
    \end{split}
\end{gather}
\endgroup
Recall, in the quenched case, the initial positions $\{\bar x_i\}$ are arranged in such a way that in the thermodynamic limit, they correspond to a well-defined macroscopic mass density profile $\varphi_q(\bar x).$ This mass distribution function is in general defined separately for right to quasiparticle $Y(t)$, denoted as $\varphi_{q,r}(\bar x)$, left to quasiparticle $X(t)$, denoted as  $\varphi_{q,\ell}(\bar x)$ and between the quasiparticles $X(t), Y(t)$, denoted as $\varphi_{q,m}(\bar x)$. One can simplify the distribution in Eq. \eqref{eq:2quasiparticle_separated_prob1_app} by rewriting it as an integral
over $\varphi_{q,r/\ell/m}(\bar x).$ First we write the product over $k$ inside the integral in Eq. \eqref{eq:2quasiparticle_separated_prob1_app} as exponential of sum over $k$ and then approximating this sum by an
integral over $\bar x$ with density $\varphi_{q,r/\ell/m}(\bar x).$ Here we use the approximation of the summation $\sum_k\mathfrak{f}(\bar{x}_k) \approx \int d\bar{x}~\varphi_q(\bar{x})~\mathfrak{f}(\bar{x})$ with the mass density $\varphi_q(\bar{x})$. With this approximation Eq. \eqref{eq:2quasiparticle_separated_prob1_app} can be written as 
\begingroup\makeatletter\def\f@size{11}\check@mathfonts
\def\maketag@@@#1{\hbox{\m@th\large\normalfont#1}}%
\begin{gather}
\begin{split}
    {\cal P}_r(m, n,  t)&=\frac{1}{4\pi^2}\int_{-\infty}^{\infty}dq_1\int_{-\infty}^{\infty}dq_2 \exp\Bigg[\int_{y_0}^{\infty}d\bar x\ \varphi_{q,r}(\bar x)\ln\Big[1+(e^{-\mc{i} a(q_1+q_2)}-1)g_<(v_0t,t|\bar x,0)\\&
    ~~~~~~~~~~~~~~~~~~~~~~~~~~~~~~~~~~+(e^{-\mc{i} aq_2}-1)g_m(v_0t,y_0+u_0t,t|\bar x,0)\Big]\Bigg]e^{\mc{i} an(q_1+q_2)}e^{\mc{i} amq_2}, \\
    {\cal P}_{m}(s,r,t)&=\frac{1}{4\pi^2}\int_{-\infty}^{\infty}dq_1\int_{-\infty}^{\infty}dq_2 \exp\Bigg[\int^{y_0}_0d\bar x~ \varphi_m(\bar x) \ln\Big[1+(e^{-\mc{i} aq_1}-1)g_{<}(y_0+u_0t,t|\bar x,0)\nonumber\\&
    ~~~~~~~~~~~~~~~~~~~~~~~~~~~~~~~~~~+(e^{-\mc{i} aq_2}-1)g_>(v_0t,t|\bar x,0)\Big]\Bigg]e^{\mc{i} a s q_1} e^{\mc{i} a r q_2},\\
    {\cal P}_{\ell}(\mu, \nu,  t)&=\frac{1}{4\pi^2}\int_{-\infty}^{\infty}dq_1\int_{-\infty}^{\infty}dq_2 \exp\Bigg[\int_{-\infty}^{0}d\bar x\ \varphi_{q,l}(\bar x)\ln\Big[1+(e^{-\mc{i} a(q_1+q_2)}-1)g_>(y_0+u_0t,t|\bar x,0)\\&
    ~~~~~~~~~~~~~~~~~~~~~~~~~~~~~~~~~~+(e^{-\mc{i} aq_1}-1)g_m(v_0t,y_0+u_0t,t|\bar x,0)\Big]\Bigg]e^{\mc{i} a\nu(q_1+q_2)}e^{\mc{i} a\mu q_2}.
\end{split}
\end{gather}
\endgroup
Once we know ${\cal{P}}(m,n,s,r,\mu,\nu,t)$ following the expressions of ${\cal P}_r(m, n,  t), {\cal P}_{m}(s,r,t), {\cal P}_{\ell}(\mu, \nu,  t),$ the moment-generating function of the $X(t)$ and $Y(t)$ can be computed as
\begin{equation}\label{eq:supp_2quasiparticle_separated_quenched_MGF1}
    {\cal{Z}}(q_1,q_2,t)=e^{-\mc{i} q_1v_0t} e^{-\mc{i} q_2(Y_0+u_0t)} ~{\cal{Z}}_r(q_1,q_2,t) ~{\cal{Z}}_{\ell}(-q_1,-q_2,t)~ {\cal{Z}}_m(q_1,q_2,t),
\end{equation}
with 
\begin{equation}
    \begin{split}
        &{\cal Z}_{r,{\ell}}(q_1,q_2,t)=\sum_{n,m}e^{-\mc{i} an(q_1+q_2)} ~ e^{-\mc{i} am q_2} ~ {\cal P}_{r,{\ell}}(m,n,t),\\&{\cal Z}_m(q_1,q_2,t)=\sum_{s,r} e^{-\mc{i} a s q_1} e^{\mc{i} arq_2} {\cal{P}}_m(s,r,t).
    \end{split}
\end{equation}
We substitute the probability distributions in Eq. \eqref{eq:supp_2quasiparticle_separated_quenched_MGF1}, expand the factor $e^{-\mc{i} q_{1,2}a}$ to quadratic orders in $q_{1,2}a$, as done in the annealed case, and then take the limit of large $N$. We get
\begingroup\makeatletter\def\f@size{10}\check@mathfonts
\def\maketag@@@#1{\hbox{\m@th\large\normalfont#1}}%
\begin{gather}\label{eq:quasiparticle2_separated_quenched_Z}
\begin{split}
{\cal Z}(q_1, q_2, t)\approx \exp \Bigg[&
-\mc{i} q_1 \Big( v_0 t + a (p_{r\ell} - p_{\ell r} - p_{\ell m}+p_{m\ell}) \Big)
\\&-\mc{i} q_2 \Big(Y_0+u_0 t + a (p_{r\ell} + p_{rm} - p_{\ell r}-p_{mr}) \Big)\\
&-\frac{q_1^2}{2} a^2 \big( p_{r\ell} + p_{\ell m} + p_{\ell r}+p_{m\ell}-w_{r\ell}- w_{\ell m}-w_{\ell r}-w_{m\ell}-2w_{\ell r m}\big)\\&
-q_1 q_2 a^2 \big( p_{r\ell} + p_{\ell r} -w_{r\ell}-w_{\ell r}-w_{r\ell m}-w_{\ell rm}+w_{m\ell r}\big)\\&
-\frac{q_2^2}{2} a^2 \big( p_{r\ell} + p_{rm} + p_{\ell r}+p_{mr} -w_{r\ell}- w_{rm} - w_{\ell r}-w_{mr}-2w_{r\ell m}\big)
\Bigg],
\end{split}
\end{gather}
\endgroup
where, $p_{\ell r}(t),p_{\ell m}(t),p_{r\ell}(t),p_{rm}(t),p_{m\ell}(t)$ and $p_{mr}(t)$ are the same as in Eqs. \eqref{eq:plR plM prL prM 2T at 0} and \eqref{eq:pmR pmL}, except now the distribution
functions $\varphi_{r/\ell/m}(\bar x)$ are replaced by  $\varphi_{q,r/\ell/m}(\bar x)$ and
\begin{align}\label{eq:2quasiparticle_origin_quenched_qprobs}
    \begin{split}
        &w_{\ell r}(t)=\int_{-\infty}^{0}d{\bar{x}}~g_{>}^2(y_0+u_0t,t|\bar{x},0)\varphi_{q,\ell}({\bar{x}}),\cr 
        &w_{\ell m}(t)=\int_{-\infty}^{0}d{\bar{x}}~g_{m}^2(v_0t,y_0+u_0t,t|\bar{x},0)\varphi_{q,\ell}({\bar{x}}),\\
        &w_{r\ell}(t)=\int_{y_0}^{\infty}d{\bar{x}}~g_{<}^2(v_0t,t|\bar{x},0)\varphi_{q,r}({\bar{x}}),\cr 
        &w_{rm}(t)=\int_{y_0}^{\infty}d{\bar{x}}~g_{m}^2(v_0t,y_0+u_0t,t|\bar{x},0)\varphi_{q,r}({\bar{x}}),\\
        &w_{m r}(t)=\int_{0}^{y_0}d{\bar{x}}~g_{>}^2(y_0+u_0t,t|\bar{x},0)\varphi_{q,m}({\bar{x}}),~\cr 
        &w_{m\ell}(t)=\int_{0}^{y_0}d{\bar{x}}~g_{<}^2(v_0t,t|\bar{x},0)\varphi_{q,m}({\bar{x}})\\
        &w_{r\ell m}(t)=\int_{y_0}^{\infty}d{\bar{x}}~g_{<}(v_0t,t|\bar{x},0)~g_{m}(v_0t,y_0+u_0t,t|\bar{x},0)~\varphi_{q,r}({\bar{x}}),\\
        &w_{m\ell r}(t)=\int_{0}^{y_0}d{\bar{x}}~g_{<}(v_0t,t|\bar{x},0)~g_{>}(y_0+u_0t,t|\bar{x},0)~\varphi_{q,m}({\bar{x}}),\\
        &w_{\ell rm}(t)=\int_{-\infty}^{0}d{\bar{x} }~g_{>}(y_0+u_0t,t|\bar{x},0)~g_{m}(v_0t,y_0+u_0t,t|\bar{x},0)~\varphi_{q,\ell}({\bar{x}}).
    \end{split}
\end{align}
Taking derivatives of $-\ln {\cal Z}(q_1, q_2, t)$ with respect to $q_1$ and $q_2$, one can compute the mean, variance, and covariance of the positions of the quasiparticles $X(t)$ and $Y(t)$ at time $t$. For the mean positions, we get
\begin{align}\label{eq:2quasiparticle_separated_quenched_mean}
    &\langle X(t) \rangle =\Big( v_0 t + a \big[p_{r\ell}(t) - p_{\ell r} (t)- p_{\ell m}(t)+p_{m\ell}(t)\big] \Big),\nonumber\\& \langle Y(t) \rangle =\Big( Y_0+u_0 t + a \big[p_{r\ell}(t) + p_{rm}(t) - p_{\ell r}(t)+p_{mr}(t)\big] \Big).
\end{align}
Similarly, the variances and covariance of the position $X(t)$ and $Y(t)$ at time $t$ become
\begin{align}
\label{eq:2quasiparticle_separated_quenched_var_covar}
\begin{split}
    \langle X^2(t)\rangle_c&=a^2 \big( p_{r\ell} (t)+ p_{\ell m} (t)+ p_{\ell r}(t)+p_{m\ell}(t)\cr 
    &~~~~~~~~~~-w_{r\ell}(t)- w_{\ell m}(t)-w_{\ell r}(t)-w_{m\ell}(t)-2w_{\ell r m}(t)\big),\\
    \langle Y^2(t)\rangle_c&= a^2 \big( p_{r\ell}(t) + p_{rm}(t) + p_{\ell r(t)}+p_{mr}(t)\cr 
    &~~~~~~~~~~-w_{r\ell}(t)- w_{rm}(t) - w_{\ell r}(t)-w_{mr}(t)-2w_{r\ell m}(t)\big),\\
    \langle X(t)Y(t) \rangle_c&=a^2 \big( p_{r\ell} (t)+ p_{\ell r}(t) -w_{r\ell}(t)-w_{\ell r}(t)+w_{m\ell r}(t)-w_{r\ell m}(t)-w_{\ell rm}(t)\big).
    \end{split}
\end{align}
It is important to note that these expressions hold for the general initial distribution of hard rods. Noting from Eqs. \eqref{eq:2quasiparticle_separated_quenched_mean} - \eqref{eq:2quasiparticle_separated_quenched_var_covar}, the mean of both positions $X(t)$ and $Y(t)$ coincide with the annealed initial condition; however, the variances and the covariance of the displacements are smaller in the quenched case than in the annealed case.
In the limit where the initial condition corresponds to zero separation between the quasiparticles ({\it i.e.} $y_0 \to 0$), the mean of the position $X(t)$ and $Y(t)$ at time $t$ becomes
\begin{align}\label{eq:2quasiparticle_origin_quenched_mean_app}
    \langle X(t) \rangle =\Big( v_0 t + a \big[p_{r\ell}(t) - p_{\ell r} (t)- p_{\ell m}(t)\big] \Big), \langle Y(t) \rangle =\Big( u_0 t + a \big[p_{r\ell}(t) + p_{rm}(t) - p_{\ell r}(t)\big] \Big),
\end{align}
and the variances of the position $X(t)$ and $Y(t)$ at time $t$ and the covariance of $X(t)$ and $Y(t)$ can be written as
\begin{align}\label{eq:2quasiparticle_origin_quenched_var_covar_app}
\begin{split}
    &\langle X^2(t)\rangle_c=a^2 \big( p_{r\ell} (t)+ p_{\ell m}(t) + p_{\ell r}(t)-w_{r\ell}(t)- w_{\ell m}(t)-w_{\ell r}(t)-2w_{\ell r m}(t)\big),\\
    &\langle Y^2(t)\rangle_c= a^2 \big( p_{r\ell}(t) + p_{rm} (t)+ p_{\ell r}(t) -w_{r\ell}(t)- w_{rm}(t) - w_{\ell r}(t)-2w_{r\ell m}(t)\big),\\
    &\langle X(t)Y(t) \rangle_c=a^2 \big( p_{r\ell}(t) + p_{\ell r}(t) -w_{r\ell}(t)-w_{\ell r}(t)-w_{r\ell m}(t)-w_{\ell rm}(t)\big).
    \end{split}
\end{align}
In Sec. \ref{sec:two quasiparticles quenched}, following Eqs. \eqref{eq:2quasiparticle_origin_quenched_mean_app} - \eqref{eq:2quasiparticle_origin_quenched_var_covar_app}, we compute the explicit expressions for the mean, variance, and covariance assuming a uniform mass density $\varrho_0$ for the background rods, in which case the corresponding point particle density is also uniform with value $\varphi_0=\frac{\varrho_0}{1-a\varrho_0}$.

%
\section{Diagonal approximation for the diffusion kernel}
\label{diag-approx}
According to linear response theory, the linearised hydrodynamics in Eq.~\eqref{BS-HD} also describes the evolution of the dynamical correlator 
\begin{equation}
\label{1}
S(x,t;v,v') = \langle \mc{f}(x,v,t)\mc{f}(0,v',0)  \rangle - \langle \mc{f}(x,v,0) \rangle 
\langle \mc{f}(0,v',0)  \rangle, 
\end{equation}
 where
$\mc{f}(x,v,t)$  is the empirical single particle phase space density at time $t$, and the average is performed over the generalized Gibbs ensemble at infinite volume. The dynamical correlation in Fourier space is defined as  
\begin{equation}
\int dx\ \mathrm{e}^{\mc{i} kx}S(x,t;v,v') = \hat{S}(k,t;v,v'). 
\end{equation}
For a one-dimensional gas of hard rods in equilibrium with background density $\varrho_0$ and velocity distribution $\mc{h}(v)$, satisfying $\mc{h}(v)=\mc{h}(-v)$,  Lebowitz, Percus, and Sykes \cite{lebowitz1968time} computed the exact dynamical correlator of hard rods. 
The exact solution for the dynamical correlator can be written as the exponential of a generator defined by,
\begin{equation}
\hat{S}(k,t;v,v')=\bra{v}\hat{S}(k,t)\ket{v'} = \bra{v}e^{tB_k} \hat{S}(k,0) \ket{v'}, \label{eq:hatS(k,t)}
\end{equation}
where $\hat{S}(k,0;v,v')= C_k(v,v')$ with $C_k(v,v')$ being the static correlation in equilibrium. Note, here $B_k$ and $C_k$ are linear operators acting on functions in velocity space, which are given explicitly by \cite{lebowitz1968time}
\begin{align}
C_k(v,v') &= \varrho_0 \Big[\mc{h}(v)\,\delta(v - v') 
+ \big(k^2(\alpha(k)^2 + \beta(k)^2)^{-1} -1\big) \mc{h}(v)\mc{h}(v')\Big], \\
B_k(v,v') &= \big[\mc{i} \beta(k)v - \alpha(k) \mu(v) \big]\delta(v - v') 
+ \alpha(k)\,\mc{h}(v)|v - v'| \notag \\
&\quad - \mc{i} (\beta(k) -k)\,\mc{h}(v)v' 
- \mc{i} \big[\beta(k) - k^{-1}(\alpha(k)^2 + \beta(k)^2)\big] v \mc{h}(v),
\end{align}
where $\varrho_0$ is the mass density of rods, related to the point particle density by $\varphi_0=\varrho_0/(1-a \varrho_0)$. Furthermore, 
\begin{align}
    \alpha(k) = \varphi_0(1- \cos(ak)),~~~ \beta(k) = k + \varphi_0\sin(ak),~~~ \mu(v) = \int dv'|v-v'| \mc{h}(v').
\end{align}

Our interest is in the large-scale behavior of the dynamical correlator. For this purpose, we expand the operators up to quadratic order in $k$
\begin{align}\label{eq:expansion_B_and_C}
    B_k \simeq \mc{i} kA- \tfrac{1}{2}Dk^2,~~~~C_k(v,v') \simeq C_0(v,v') + \tfrac{1}{2}k^2 C_{[2]}(v,v'),
\end{align}
where,
\begin{eqnarray}
&& A(v,v')  = v_{\rm eff}(v) \delta(v-v') + a\varphi_0v_{\rm eff}(v) \mc{h}(v) - a\varphi_0\mc{h}(v) v',\\[1ex]
 &&D(v,v') = a^2\varphi_0\left[\delta(v-v') \mu(v) - \mc{h}(v)|v-v'|\right], \label{D(v,v')}\\[1ex]
&& C_0(v,v')  = \varrho_0 \left[\mc{h}(v) \delta(v-v') + a\varrho_0(a\varrho_0 -2)  \mc{h}(v) \mc{h}(v')\right] , \\[1ex]
&& C_{[2]}(v,v')= \tfrac{2}{3}a^3 (1+ a\varphi_0)^{-3}(1 + \tfrac{1}{4}\varphi_0) \varrho_0^2 \mc{h}(v) \mc{h}(v'),
\end{eqnarray}
with the effective velocity defined as
\begin{align}
    v_{\rm eff}(v) =(1+a \varphi_0) v = \frac{v}{1-a \varrho_0},
\end{align}
and $C_0(v,v')$ is the integral kernel of the static correlator. If we expand the Euler equations around the equilibrium distribution $\varrho_0 \mc{h}(v)$, the fluctuations evolve according to the operator $\mc{i} kA$. On general grounds,  the operator $AC_0$ is expected to be symmetric, which is verified. To exponentiate by $\mc{i} kAt$, one has to diagonalize $A$ which is achieved by $R^{-1}AR=V_{\rm eff}$ where $\bra{v}V_{\rm eff}\ket{v'}=v_{\rm eff}(v) \delta (v-v')$ and 
\begin{align}
\label{8}
\bra{v}R\ket{v'}=R(v,v') &= \delta(v-v') -a\varrho_0 \mc{h}(v), \notag \\
\bra{v}R^{-1}\ket{v'}=R^{-1}(v,v')& =  \delta(v-v') + a \varphi_0 \mc{h}(v),
\end{align} 
satisfying $ R^{-1}R = 1 = R R^{-1}$.
Then, the time-evolution operator can be written as 
\begin{align}
   \msc{J}_t= R R^{-1} e^{\mc{i}kt A} R R^{-1} &= R e^{\mc{i}ktR^{-1}AR} R^{-1} = R e^{\mc{i}kt V_{\rm eff}}R^{-1},
\end{align}
whose integral kernel, denoted by $\msc{J}_t(v,v')$, takes the following form
\begin{gather}\label{eq:kernel_A}
    \begin{split}
       \msc{J}_t(v,v') =& \delta(v-v') e^{\mc{i}kt~v_{\rm eff}(v)} - a\varrho_0 \mc{h}(v) e^{\mc{i}kt~v_{\rm eff}(v')} + a \varphi_0 \mc{h}(v) e^{\mc{i}kt~v_{\rm eff}(v)} \\& - a^2 \varrho_0 \varphi_0 \mc{h}(v) \int dw ~ \mc{h}(w) e^{\mc{i}kt~v_{\rm eff}(w)}.
    \end{split}
\end{gather}
Now, extending up to quadratic order, we add the diffusive term $D$ in the generator. Surprisingly, one finds 
\begin{align}
    R^{-1}DR=D,
\end{align}
which allows one to merely add the diffusive correction $D$ to $A$, in Eq.~\eqref{eq:expansion_B_and_C}. Hence the time-evolution operator $e^{tB_k}$, with $B_k$ given in Eq.~\eqref{eq:expansion_B_and_C}, can be written as 
\begin{align}
e^{tB_k}=RR^{-1}e^{tB_k}RR^{-1} = R~e^{\mc{i}kt~V_{\rm eff} -\frac{k^2 D t}{2}} R^{-1}.
\end{align}
Using this in Eq.~\eqref{eq:hatS(k,t)} we write 
\begin{align}
\hat{S}(k,t) = e^{tB_k}\hat{S}(k,0)= R~e^{\mc{i}kt~V_{\rm eff} -\frac{k^2 D t}{2}} R^{-1} \hat{S}(k,0), \label{eq:hatS(k,t)-2}
\end{align}
where recall $\hat{S}(k,0) \simeq C_0 + \tfrac{k^2}{2}C_{[2]}$ (see Eq.~\eqref{eq:expansion_B_and_C}). 
One can neglect the contribution from $O(k^2)$ correction ($C_{[2]}$) to the static correlator, as in real space, combined with the Gaussian kernel, this would yield a correction of order $1/t$ which is subleading.
Since the operators $V_{\rm eff}$ and $D$ do not commute, no further simplification appears to be possible in Eqs.~\eqref{eq:hatS(k,t)-2}.
Note that from Eq.~\eqref{eq:hatS(k,t)-2} or directly from Eq.~\eqref{eq:hatS(k,t)} it follows that
\begin{align}
\partial_t \hat{S}(k,t) = \left(\mc{i}k A -\frac{k^2}{2}D\right)~\hat{S}(k,t),
\label{eq:hatS(k,t)-3}
\end{align}
in agreement with the HD equation \eqref{BS-HD}, linearized around generalized equilibrium. In the Euler frame the correlation, defined as $\hat{S}^{\rm e}(k,t;v,v') = e^{-\mc{i}kAt}\hat{S}(k,t;v,v')$ evolves as 
\begin{align}
 \partial_t \hat{S}^{\rm e}(k,t) &=-\frac{k^2}{2}~e^{-\mc{i}kt A} D  e^{\mc{i}kt A}  \hat{S}^{\rm e}(k,t), \notag \\ 
 &= \left[-\frac{k^2}{2}D +\mc{i}\frac{k^3t}{2}[A,D] +O\left(k^4\right)\right]~\hat{S}^{\rm e}(k,t), \label{eq:hatS^e}
\end{align}
where $[A,D]\neq 0$ is the commutator between the two operators. 
In kinetic theory, the relaxation time approximation is very common, which amounts to keeping only the $\mu(v)$ term {\it i.e.} $D \approx D_{\rm diag}$ with 
$\bra{v}D_{\rm diag}\ket{v'}=\delta(v-v') a^2 \varphi_0 ~\mu(v)$.
The rationale for the diagonal approximation is that $\mu(v)$ results in a continuous spectrum, while the off-diagonal term has a pure point spectrum. In a similar context \cite{lukkarinen2008anomalous}, for a wave kinetic equation in one dimension, the second-order resolvent expansion was used to confirm the relaxation time approximation. 

With the above diagonal approximation for $D$ and keeping terms up to quadratic order in $k$,  Eq.~\eqref{eq:hatS^e} becomes 
\begin{align}
 \partial_t \hat{S}^{\rm e}(k,t) \approx -\frac{k^2}{2}D_{\rm diag}\hat{S}^{\rm e}(k,0). \label{eq:hatS(k,t)-4} 
\end{align}
Ferrari and Olla \cite{ferrari2023macroscopic} establish a Langevin equation for the empirical density on one-particle phase space valid up to the diffusive time scale. The covariance computed from this equation agrees with \eqref{eq:hatS(k,t)-4}.
\end{appendix}

\bibliography{references}
\nolinenumbers
\end{document}